\documentclass[prx,twocolumn,amsmath,amssymb,superscriptaddress,floatfix,nofootinbib,aps]{revtex4-2}
\usepackage{amsmath}
\usepackage{amsfonts}
\usepackage{graphicx}
\usepackage{dcolumn}
\usepackage{bm}
\usepackage{amsthm}
\usepackage{algorithm}
\usepackage{algpseudocode}
\usepackage[cmyk]{xcolor}
\usepackage[colorlinks,bookmarks=false,citecolor=magenta,linkcolor=magenta,urlcolor=magenta]{hyperref}
\usepackage{braket}
\usepackage{orcidlink}

\begin{document}
\preprint{APS/123-QED}
\title{Nonlocal nonstabilizerness for slightly entangled quantum many-body states}

\author{Lei-Yi-Nan Liu \orcidlink{0009-0005-3072-3650}}
 \affiliation{%
 School of Physics, Beihang University, Beijing 100191, China
}

\author{Jian Cui \orcidlink{0000-0001-6643-7625}}
 \email{jiancui@buaa.edu.cn}
 \affiliation{%
 School of Physics, Beihang University, Beijing 100191, China
}

\date{\today}

\begin{abstract}
Nonlocal nonstabilizerness quantifies the irreducible magic resource encoded in bipartite entanglement, but its evaluation is generally hindered by a highly nonconvex optimization over local unitary transformations. 
Here we propose a Schmidt-reference-state framework that replaces this optimization by a direct construction from the sorted Schmidt spectrum. 
We conjecture that the nonlocal stabilizer R\'enyi entropy (SRE) is given by the SRE of the corresponding reference state, and support this conjecture through analytical and numerical evidences. 
Our framework makes nonlocal nonstabilizerness efficiently accessible for weakly entangled many-body states whenever the entanglement spectrum is available. 
Applying it to Haar-random states, critical spin chains, and PXP dynamics, we show that nonlocal SRE captures nonstabilizer structures in the entanglement spectrum that are invisible to conventional entanglement measures. 
Our results establish entanglement spectra as a powerful window into irreducible nonstabilizer correlations, opening a broadly applicable route to studying nonlocal magic resources of quantum many-body systems in and out of equilibrium.
\end{abstract}

\maketitle
\section{introduction}
Entanglement stands for a core concept across quantum information and quantum many-body physics. It describes nonlocal correlations and underpins a broad range of key phenomena~\cite{Horodecki2009}, including quantum phase transitions~\cite{Osborne2002}, topological order~\cite{Tantivasadakarn2024,Chen2010,XiaoGang2019,Broholm2020}, thermalization~\cite{Abanin2019,Kaufman2016,Brenes2020}, and tensor-network simulability~\cite{Vidal2003,Sauerwein2019,Cirac2021}. 
However, entanglement alone does not fully determine the computational complexity of quantum states, since highly entangled stabilizer states can still be efficiently simulated classically~\cite{Gottesman1998}. 
Nonstabilizerness, also known as the magic resource, serves as a complementary quantum resource distinct from entanglement. It quantifies the deviation from the stabilizer formalism and supplies the extra resource necessary to realize universal quantum computation~\cite{Bravyi2005,Howard2017,Veitch2012}. 
This complementarity has motivated the development of stabilizer Rényi entropies (SREs)~\cite{Lorenzo2022}, a set of computable quantifiers for nonstabilizerness that have been widely applied to equilibrium quantum phases~\cite{Haug2023MPS,Tarabunga2023,Tarabunga2024,Tarabunga2024critical,Hallam2026}, quantum many-body dynamics~\cite{Falcao2025,Russomanno2025,Grabarits2026,Odavic2025,Smith2025}, and quantum circuits~\cite{Zhenyu2026,Maity2026,Turkeshi2025,Szombathy2025,Haug2025,Niroula2024}.

A fundamental question concerns the interrelation between entanglement and nonstabilizerness. 
The notion of nonlocal magic, also called nonlocal nonstabilizerness, addresses this question by extracting the component of nonstabilizerness invariant under elimination via local unitary transformations~\cite{Qian2025,Cao2025}. 
For a bipartite pure state, it is defined by minimizing the SRE over local unitaries acting separately on the two subsystems. This construction removes purely local contributions and captures the irreducible nonstabilizerness associated with the bipartite entanglement structure. 
Despite its conceptual importance, the nonlocal magic resource is extremely difficult to evaluate, since its definition involves a highly nonconvex optimization over exponentially large local-unitary manifolds. 
Direct minimization is therefore restricted to very small Hilbert spaces, and analytical results are known only in special cases, such as two-qubit states~\cite{Qian2025,Cao2025}, certain prime-dimensional qudit systems~\cite{Giorgio2026}, and fermionic Gaussian states~\cite{Daniele2026,Mario2026}. 
In the latter case, restricting the minimization to local Gaussian unitaries yields a closed-form expression in terms of the reduced Majorana covariance matrix, but this does not solve the general local-unitary problem for interacting many-body states. 
Recent work has shown that, for bipartite pure states, the nonlocal magic resource is completely determined by the nonzero Schmidt spectrum, establishing a fundamental link between entanglement structure and nonlocal nonstabilizerness~\cite{Xiao2026,Gianpaolo2026}. 
However, this spectral characterization does not by itself provide any efficient evaluation method, because the nonlocal SRE depends nontrivially on the Schmidt coefficients. Thus, even after reducing the problem to the entanglement spectrum, the practical computation of nonlocal nonstabilizerness remains challenging.

In this work, we conjecture an analytical expression for the nonlocal SRE directly in terms of the Schmidt spectrum. 
By assigning sorted Schmidt coefficients to a canonical reference state, the nonlocal SRE coincides with the SRE of the resulting state, thereby circumventing the original nonconvex local-unitary optimization and enabling direct spectral calculation. 
We provide analytical validation by proving that the reference state satisfies the requisite first-order optimality condition. Furthermore, we benchmark our conjecture against direct manifold optimization over a broad family of rank-4 states and representative many-body states, including the ground states of the Ising and XXZ models~\cite{Smart2024,James2016,Jiang2019}.
The excellent agreement across feasible system sizes demonstrates that this construction enables efficient evaluation of nonlocal nonstabilizerness, especially for weakly entangled many-body states including tenser-network states and ground states of gapped local Hamiltonians. 
We further compare our nonlocal magic resource (NL) with the fermionic nonlocal magic resource (FNL)~\cite{Daniele2026,Mario2026}. 
While FNL admits efficient computation for Gaussian states, it minimizes exclusively over local Gaussian unitaries, thereby yielding a restrictive upper bound. In contrast, our reference-state construction targets the fully local-unitary-minimized nonstabilizerness and captures fine-grained features of many-body entanglement-spectrum structures beyond Gaussian covariance information. We prove that the reference-state SRE is upper-bounded by FNL, demonstrating its capability as a sharper spectral estimator for nonlocal nonstabilizerness. We subsequently apply our framework to Ising and XXZ spin chains, as well as to PXP dynamics in Rydberg atom arrays. These applications reveal three distinctive properties of nonlocal SRE: it exhibits logarithmic critical scaling, varies throughout the interacting XXZ critical phase even at a fixed central charge, and unambiguously differentiates entanglement growth from the generation of irreducible nonstabilizerness in scarred quantum dynamics.

The paper is organized as follows. 
In Sec.~\ref{sec1}, we introduce the basic concept of SRE, define the nonlocal SRE through local-unitary minimization, and present our conjectured Schmidt reference state and the analytical expression of nonlocal SRE based on it. 
We also discuss the truncation of small Schmidt coefficients for nonlocal SRE evaluation. 
In Sec.~\ref{sec2}, we develop a manifold-optimization method to verify our conjecture, finding excellent agreement with the conjectured analytical expression for a broad class of rank-4 spectra. 
We further analyze the rank-4 case and derive several analytical properties of the nonlocal SRE.
In Sec.~\ref{sec3}, we apply our framework to Haar-random states and representative many-body systems, including the ground states of  Ising and XXZ models, as well as the time-evolved states of PXP model. By comparing with the fermionic nonlocal magic resource and conventional entanglement entropies, we show that nonlocal SRE captures finer entanglement-spectrum structures both in and out of equilibrium.
Finally, in Sec.~\ref{sec4}, we summarize our results and discuss future applications of nonlocal SRE expression to tensor network states, higher dimensional systems, many-body localized systems, topologically ordered systems and quantum circuits.

\section{Nonlocal SRE}\label{sec1}
We begin by introducing the Stabilizer R\'enyi Entropy (SRE), a widely adopted measure of magic resources.
The $\alpha$-stabilizer R\'enyi entropy ($\alpha$-SRE) for a $n$-qubit pure state $\ket{\psi}$ is defined as 
\begin{equation}
    M_{\alpha}(\ket{\psi}):=\frac{1}{1-\alpha}\log_2\sum_{P\in\mathcal{P}_n}\Xi_P^{\alpha}(\ket{\psi})-\log_2 d \label{definition}
\end{equation}
where $\mathcal{P}_n$ is the group of all $n$-qubit Pauli strings with $+1$ phase, and 
$\Xi_P(\ket{\psi}):=d^{-1}\bra{\psi}P\ket{\psi}^2$ with $d=2^n$ the dimension of $n$-qubit Hilbert 
space \cite{Lorenzo2022}. 
Now we partition the state $\ket{\psi}$ into $A$ and $B$ subsystems with $n_A$ and $n_B$ qubits respectively, we can perform the Schimidt decomposition as 
\begin{equation}
    \ket{\psi}_{AB} = \sum_{i=0}^{\chi-1}\sqrt{\lambda_i}\ket{a_i}_A\ket{b_i}_B
\end{equation}
where $\chi$ denotes the Schmidt rank, i.e., the number of nonzero Schmidt coefficients. 
The Schmidt coefficients satisfy $\sum_i\lambda_i=1$ and the Schmidt basis states are orthonormal $\braket{a_i|a_j}=\braket{b_i|b_j}=\delta_{ij}$. 
Without loss of generality, the Schmidt coefficients are arranged in descending order as 
$\lambda_0\ge\lambda_1\ge...\ge\lambda_{\chi-1}>0$. 

The nonlocal $\alpha$-SRE $M_{\alpha}^{\mathrm{NL}}$ for $\ket{\psi}_{AB}$ is defined as
\begin{equation}
M_{\alpha}^{\mathrm{NL}}(\ket\psi_{AB}):=\min_{U_A\otimes U_B} M_{\alpha}(U_A\otimes U_B\ket{\psi}_{AB}),
\end{equation}
where $U_A$ and $U_B$ are unitary operators acting on subsystems $A$ and $B$, respectively \cite{Qian2025}.
We now introduce a Schmidt reference state 
\begin{equation}
    \ket{\tilde{\psi}}_{\tilde{A}\tilde{B}}=\sum_{i=0}^{\chi-1}\sqrt{\lambda_i}\ket{c_i}_{\tilde{A}}\ket{c_i}_{\tilde{B}}
    \label{eq4}
\end{equation}
where $\ket{c_i}$ denotes the computational basis state labelled by $i$. 
Here $\tilde A$ and $\tilde B$ are auxiliary effective subsystems associated with the Schmidt supports of the original subsystems, whose Hilbert spaces are chosen to contain $\lceil \log_2 \chi\rceil$ qubits. 
They do not represent specific physical subsystems, but provide a canonical computational-basis encoding of the ordered Schmidt coefficients, so that $\ket{\tilde{\psi}}$ has exactly the same Schmidt spectrum as the original state $\ket{\psi}$. 
Recently, Huang \emph{et al.} proved that the bipartite nonlocal magic resource depends solely on the nonzero Schmidt spectrum of a bipartite pure state \cite{Xiao2026}. The reference state defined in this work coincides with the canonical encoding state introduced in Ref.~\cite{Xiao2026} and also the Schmidt-gauged representative introduced in Refs.~\cite{Gianpaolo2026,Fabio2026}. Consequently, we have 
\begin{equation}
    M_{\alpha}^{\mathrm{NL}}(\ket{\psi})=M_{\alpha}^{\mathrm{NL}}(\ket{\tilde{\psi}})\le M_{\alpha}(\ket{\tilde{\psi}}).
\end{equation}
Here the SRE of reference state offers a theoretical upper bound of nonlocal SRE of the original state $\ket{\psi}$ \cite{Cao2025,Xiao2026}. 
Motivated by extensive numerical evidence presented below, 
we conjecture that the above bound is saturated for generic quantum states, i.e., 
\begin{equation}
    M_{\alpha}^{\mathrm{NL,ref}}(\boldsymbol{\lambda}):=M_{\alpha}(\ket{\tilde{\psi}})=M_{\alpha}^{\mathrm{NL}}(\ket{\psi}), \label{conjecture}
\end{equation}
where $\boldsymbol{\lambda} = \{\lambda_i\}$ is the entangement spectrum of state $\ket{\psi}$ and $\ket{\tilde{\psi}}$. 
Namely, the nonlocal SRE of a bipartite pure state is exactly given by the SRE of its reference state. 

To support this conjecture, we prove that the reference state is a stationary point of the SRE under arbitrary local unitary transformations. More explicitly, for any local Hermitian generator
\begin{equation}
    H=H_{\tilde{A}}\otimes \mathbb{I}_{\tilde{B}}
    +\mathbb{I}_{\tilde{A}}\otimes H_{\tilde{B}},
\end{equation}
and for any R\'enyi index $\alpha$, we prove that
\begin{equation}
    \left.
    \frac{\partial}{\partial t}
    M_{\alpha}\left(e^{-iHt}\ket{\tilde{\psi}}\right)
    \right|_{t=0}
    =0 .\label{eq8}
\end{equation}
Here $e^{-iHt}$ generates an arbitrary infinitesimal local unitary transformation of the form
$U_{\tilde{A}}\otimes U_{\tilde{B}}$ acting on the reference state $\ket{\tilde{\psi}}$.

This result provides nontrivial evidence for our conjecture. Since the local unitary manifold
$\mathrm{SU}(d_{\tilde{A}})\times \mathrm{SU}(d_{\tilde{B}})$ is compact, the SRE must attain its minimum and maximum on this manifold. At any interior extremum, the first-order variation along every local-unitary direction must vanish. The above result shows that the reference state satisfies precisely this necessary condition. It is a critical point of the SRE landscape under local unitary transformations. See Appendix~\ref{app1} for proof and further discussion. 

Moreover, the reference state is not expected to be a maximizer of the SRE. One can easily find local unitary transformations that increase the SRE relative to the reference state value. Therefore, the stationary character of the reference state is more naturally interpreted as evidence that it is a candidate minimizer rather than a maximizer. In addition, following the descending-order prescription for the entanglement spectrum, the reference state appears to give the smallest SRE among the stationary points we have identified~\cite{Cao2025}. These observations strongly suggest that the SRE of the descending-order reference state gives the nonlocal SRE. This expectation is further supported by our numerical optimization results, which show good agreement between the optimized nonlocal SRE and the reference state SRE up to numerical precision. 
In the following, we mainly focus on the 2-SRE, namely the case $\alpha=2$, unless otherwise specified, since it is the lowest-order SRE that defines a legitimate magic monotone among the family with $\alpha\ge 2$~\cite{Haug2023monotone,Lorenzo2024}.

For the Schmidt-rank-2 case ($\chi=2$), the nonlocal SRE can be derived analytically \cite{Giorgio2026,Xiao2026,Cao2025,Qian2025,Fabio2026}. Parameterizing the Schmidt spectrum as $\boldsymbol{\lambda}=(\cos^2\theta,\sin^2\theta)$, 
one obtains 
\begin{equation}
    M_{2}^{\mathrm{NL}}(\boldsymbol{\lambda})=M_{2}^{\mathrm{NL}}(\theta)=\log_2\left(\frac{8}{7+\cos8\theta}\right).
\end{equation}
For higher Schmidt ranks, our conjecture implies that the nonlocal SRE is given by 
\begin{eqnarray}
    &M_{2}^{\mathrm{NL,ref}}(\boldsymbol{\lambda})=-\log_2\Bigg(
    \displaystyle\sum_{i_1,i_2,i_3,i_4=0}^{2^n-1}\sqrt{\lambda_{i_1}\lambda_{i_2}\lambda_{i_3}\lambda_{i_4}}\times\nonumber\\
    &\sqrt{\lambda_{i_3\oplus i_2\oplus i_1}\lambda_{i_4\oplus i_2\oplus i_1}\lambda_{i_1\oplus i_3\oplus i_4}\lambda_{i_2\oplus i_3\oplus i_4}}
    \Bigg), \label{MnlEq}
\end{eqnarray}
where $\oplus$ denotes the bitwise XOR operation. 
This equation was first derived in Ref.~\cite{Poetri2024} in a different context and was 
subsequently discussed in Ref.~\cite{Cao2025}. 
As emphasized in Ref.~\cite{Cao2025}, the value of Eq.~\eqref{MnlEq} depends on how the Schmidt coefficients are assigned to the computational-basis states. Among all possible assignments, the descending ordering $\lambda_0\ge\lambda_1...\ge \lambda_{2^n-1}$ 
yields the smallest value of Eq.~\eqref{MnlEq}, and therefore provides the tightest upper bound on the nonlocal SRE. 
More recently, Refs.~\cite{Gianpaolo2026, Fabio2026} interpreted Eq.~\eqref{MnlEq} in terms of the Walsh-Hadamard autocorrelations of the entanglement spectrum, providing a more convenient framework for theoretical analysis. 
To evaluate Eq.~\eqref{MnlEq}, the Schmidt spectrum must be embedded into a Hilbert space of dimension $2^n$. When $\chi$ is not a power of 2, the Schmidt spectrum is padded with zeros until its dimension reaches the smallest power of 2 satisfying $2^{n-1}<\chi\le2^n$. 

Although Eq.~\eqref{MnlEq} provides a direct way to evaluate the conjectured nonlocal SRE from the Schmidt spectrum, 
its computational cost still grows exponentially with the effective number of qubits $n$. 
For small Schmidt spectra this is not a serious limitation, and Eq.~\eqref{MnlEq} can be evaluated exactly. 
However, for many-body systems the Schmidt rank typically grows exponentially with the subsystem size, 
making a direct evaluation increasingly difficult. 
This difficulty persists even if the entanglement spectrum is generated efficiently, for example from a free-fermion correlation matrix as we will discuss later, because the resulting many-body Schmidt spectrum still contains exponentially many entries. 
Thankfully, in practice, the entanglement spectrum often decays rapidly. 
This property motivates a controlled truncation scheme in which only the largest Schmidt probabilities are retained. 
Let the full Schmidt spectrum be $\{ \lambda_i \}$, and let $D$ denote the set of discarded Schmidt values with the total weight $\eta=\sum_{i\in D}\lambda_i$. 
We proved that the truncation error is controlled by the total discarded weight $\eta$. 
Specifically, we obtain 
\begin{equation}
    \left|
    M_\alpha(\ket{\tilde{\psi}})-
    M_\alpha(\ket{\tilde{\psi}_\mathrm{trunc}})
    \right|
    \lesssim
    \frac{2\alpha\sqrt{2}}{|\alpha-1|\ln2}
    \frac{\sqrt{\eta}}{\zeta_\alpha(\ket{\tilde{\psi}})},
    \label{eq:M_trunc_bound_main}
\end{equation}
provided that the change of $\zeta_\alpha(\ket{\tilde{\psi}})$, due to spectrum truncation, is small compared with $\zeta_\alpha(\ket{\tilde{\psi}})$ itself, 
where $\zeta_\alpha(\ket{\tilde{\psi}}):=\sum_{P\in\mathcal{P}_n}\Xi_P^{\alpha}(\ket{\tilde{\psi}})$, and $\ket{\tilde{\psi}_\mathrm{trunc}}$ is the reference state with a truncated spectrum. 
Therefore, the approximation is controlled by the accumulated discarded weight $\eta$. 
In the critical systems considered below, where the nonlocal SRE grows at most logarithmically with subsystem size $\ell$,
$M_\alpha^{\mathrm{NL}}(\ell)=O(\log\ell)$, 
one has $\zeta_\alpha=2^{(1-\alpha)M_\alpha^{\mathrm{NL}}}=\ell^{-O(1)}$. 
Thus Eq.~\eqref{eq:M_trunc_bound_main} implies
\begin{equation}
    \Delta M_\alpha^{\mathrm{NL}}
    \le \mathrm{poly}(\ell)\sqrt{\eta},
\end{equation}
where $\Delta M_\alpha^{\mathrm{NL}}$ is the truncation error. 
This justifies using a truncated entanglement spectrum in the following calculations, 
provided that the accumulated discarded weight $\eta$ is explicitly monitored and kept sufficiently small. 
The proof of this bound is given in Appendix~\ref{app:truncation_error}.

\section{Numerical optimization}\label{sec2}

\begin{figure}
    \includegraphics[width=\linewidth]{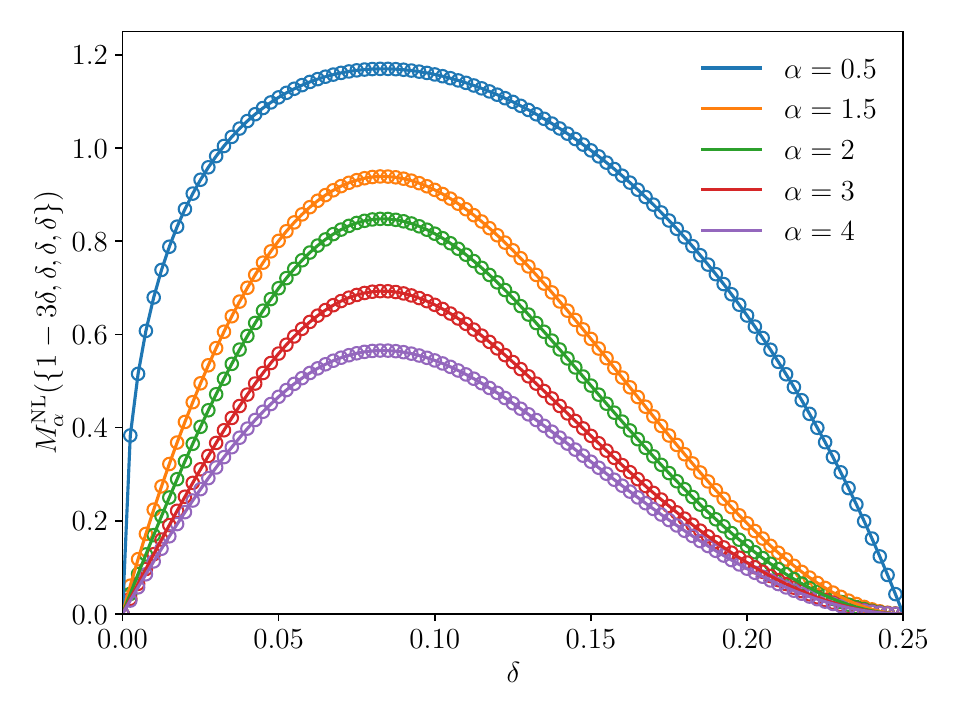}
    \caption{
    Nonlocal SRE for the entanglement spectrum
    $\{1-3\delta,\delta,\delta,\delta\}$ at different R\'enyi indices $\alpha$.
    The solid lines denote the SRE of the corresponding reference states,
    while the open circles represent the results obtained from manifold optimization.
    \label{fig1}
    }
\end{figure}

To numerically evaluate the nonlocal SRE of a general state $\ket{\psi}$, it is sufficient to optimize over its reference state. This reduces the search space from $\mathrm{SU}(d_A)\otimes\mathrm{SU}(d_B)$ to $\mathrm{SU}(d_{\tilde A})\otimes\mathrm{SU}(d_{\tilde B})$, where
$d_{\tilde A}=d_{\tilde B}=2^{\lceil\log_2\chi\rceil}$, 
and $\chi$ denotes the Schmidt rank of $\ket{\psi}$. For slightly entangled quantum states, $\chi$ is typically much smaller than the subsystem Hilbert-space dimensions, leading to a substantial reduction in computational cost.

However, this reduction is not optimal. For example, consider a state with Schmidt rank $\chi=2^n+1$. Even after introducing the reference state, one still has to optimize over $\mathrm{SU}(2^{n+1})\otimes\mathrm{SU}(2^{n+1})$, since the reference-state Hilbert-space dimension is determined by the smallest power of 2 containing the Schmidt support. In practice, this introduces many redundant degrees of freedom. Indeed, only $\chi$ orthonormal basis vectors are required to specify the Schmidt subspace, making it unnecessary to optimize over the entire unitary group.

These $\chi$ orthonormal vectors naturally form an element of the complex Stiefel manifold, defined as 
\begin{equation}
    \mathrm{St}_{\mathbb C}(d,\chi)
    :=
    \left\{
    U\in\mathbb{C}^{d\times\chi}
    \;\middle|\;
    U^\dagger U=\mathbb I_\chi
    \right\}.
\end{equation}
Therefore, the optimization can be reformulated on $\mathrm{St}_{\mathbb C}(d,\chi)$ instead of $\mathrm{SU}(d)$. The real dimension of $\mathrm{SU}(d)$ is $d^2-1$
whereas the complex Stiefel manifold has real dimension
$2d\chi-\chi^2$. 
Consequently, when $\chi\ll d$, the number of variational degrees of freedom is dramatically reduced, leading to a substantial improvement in optimization efficiency.

Unlike $\mathrm{SU}(d)$, which admits a convenient parameterization in terms of its Lie-algebra generators, the complex Stiefel manifold does not possess an equally simple global parameterization. For this reason, rather than explicitly parameterizing the manifold, we employ Riemannian manifold optimization techniques \cite{James2016,Smart2024,Jiang2019}. In this approach, the optimization is performed directly on the manifold by projecting the Euclidean gradient onto the tangent space and subsequently retracting the updated point back onto the manifold, thereby preserving the orthonormality constraints throughout the optimization procedure. For brevity, we omit the technical details of the Riemannian optimization algorithms and refer the reader to Ref.~\cite{Jiang2019} for comprehensive discussions. 
In this paper, we use \verb|pymanopt| to implement optimization on complex Stiefel manifold \cite{James2016}. 

We first consider the $\chi=4$ case, for which the Schmidt spectrum contains three independent parameters. To showcase the agreement between the optimized nonlocal SRE and the reference-state prediction, we consider the one-parameter family of entanglement spectra $\boldsymbol{\lambda}=(1-3\delta,\delta,\delta,\delta)$, 
for which the SRE of the corresponding reference state can be evaluated analytically and for $\alpha=2$ we have 
\begin{equation}
    M_2^{\mathrm{NL,ref}}(\delta)=-\log_2\left[1-12\delta(1-4\delta)^2\right]. 
\end{equation}
In Fig.~\ref{fig1}, we compare the nonlocal SRE obtained from manifold optimization with the reference state result for several R\'enyi indices $\alpha$.
As shown in Fig.~\ref{fig1}, the numerical optimization results are indistinguishable from the reference-state predictions over the entire range $\delta\in[0,1/4]$. For all values of $\alpha$ considered, the optimized nonlocal SRE coincides with the SRE of the reference state within numerical precision. This agreement provides strong evidence that the reference state indeed realizes the global minimum of the local-unitary optimization problem, supporting the conjecture
$M_{\alpha}^{\mathrm{NL}}(\ket{\psi})=M_{\alpha}(\ket{\tilde{\psi}})$.
Moreover, the agreement persists not only for $\alpha=2$, but also for a broad range of R\'enyi indices. This suggests that the reference-state description captures a universal feature of bipartite nonlocal nonstabilizerness rather than a special property of a particular fixed $\alpha$-SRE.

\begin{figure}
    \includegraphics[width=\linewidth]{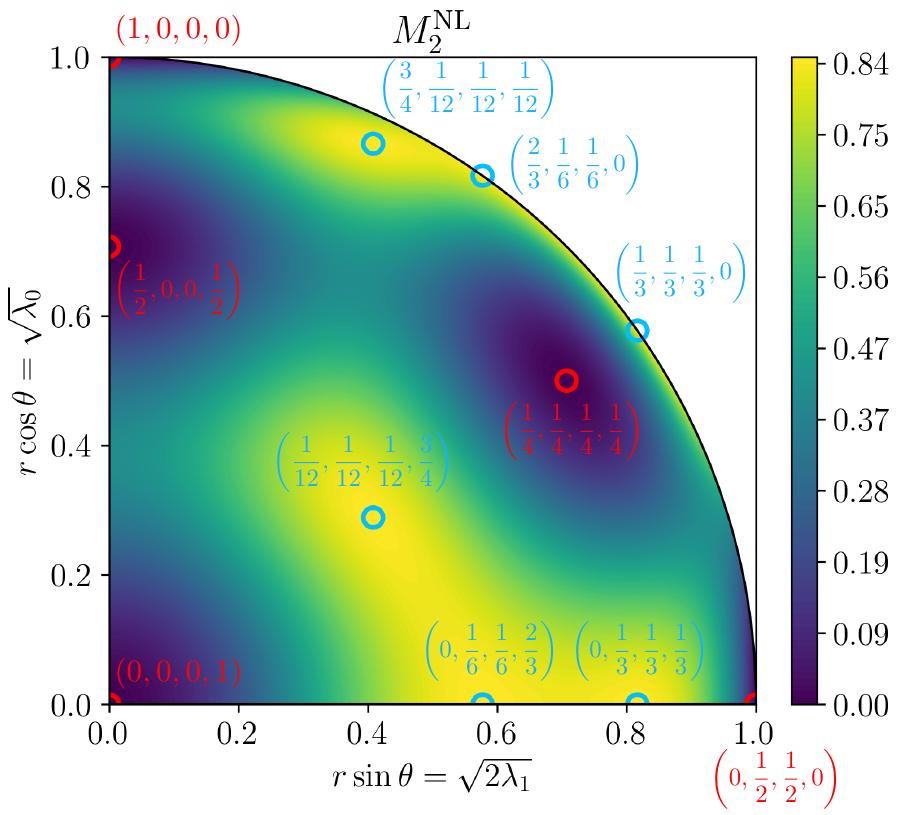}
    \caption{
    Nonlocal SRE $M_2^{\mathrm{NL}}$ 
    for rank-4 entanglement spectra, parameterized by 
    $(r,\theta,\phi=\pi/4)$ as defined in Eq.~\eqref{parameterization4}. 
    The red open circles indicate the spectra with vanishing nonlocal 
    SRE, while the blue open circles denote the global maximizers of 
    the nonlocal SRE over the space of rank-4 entanglement spectra.
    \label{fig2}
    }
\end{figure}

We then parameterize a rank-4 entanglement spectrum $\boldsymbol{\lambda}=(\lambda_0,\lambda_1,\lambda_2,\lambda_3)$ as
\begin{eqnarray}
    \lambda_0&=&r^2\cos^2\theta,\quad \lambda_1=r^2\sin^2\theta\cos^2\phi,\nonumber\\
    \lambda_2&=&r^2\sin^2\theta\sin^2\phi,\quad \lambda_3=1-r^2,
    \label{parameterization4}
\end{eqnarray}
and numerically optimize the nonlocal SRE over the three parameters $(r,\theta,\phi)$.
Our numerical results strongly support the conjecture that, for calculated rank-4 entanglement spectra, the nonlocal SRE coincides with the SRE of the corresponding reference state up to numerical precision.
Motivated by this observation, we carry out several analytical analyses of the rank-4 case based on the conjecture.
The spectra with vanishing nonlocal SRE are marked by red open circles in Fig.~\ref{fig2}. They correspond to the spectra
$(1,0,0,0)$, $(\frac12,\frac12,0,0)$, and $(\frac14,\frac14,\frac14,\frac14)$.
The spectra attaining the maximum nonlocal SRE are marked by blue open circles.
Using the analytical expression of the nonlocal SRE for the reference states, we find that there are three classes of spectra that maximize the nonlocal SRE, namely $(\frac13,\frac13,\frac13,0)$, $(\frac23,\frac16,\frac16,0)$, and $(\frac34,\frac1{12},\frac1{12},\frac1{12})$.
For all three cases, the maximum nonlocal SRE is $\log_2(9/5)\approx0.848$. 

We further find that, based on our conjecture, the nonlocal SRE for rank-4 entanglement spectra admits a remarkably simple expression in terms of the R\'enyi entropies. The $\alpha$-R\'enyi entropy is defined as
\begin{equation}
    S_{\alpha}=\frac{1}{1-\alpha}\log_2\operatorname{Tr}\left(\rho^{\alpha}\right)
    =\frac{1}{1-\alpha}\log_2\left(\sum_i\lambda^{\alpha}_i\right),
\end{equation}
where $\rho$ is the reduced density matrix. In the rank-4 case, we obtain
\begin{equation}
    2^{-M_2^{\mathrm{NL,ref}}}=7-42\cdot2^{-S_2}+28\cdot2^{-2S_2}+56\cdot2^{-2S_3}
    -48\cdot2^{-3S_4}, \label{lowrankapprox}
\end{equation}
where both $M_2^{\mathrm{NL,ref}}$ and $S_{\alpha}$ are evaluated from the same rank-4 entanglement spectrum. 
It is worth emphasizing that this relation is specific to the rank-4 case. For higher Schmidt ranks, although the nonlocal SRE is still expected to be determined by the entanglement spectrum under our conjecture, it is generally not expected to admit such a simple closed-form expression solely in terms of a finite number of low-order R\'enyi entropies.
\footnote{
For $\chi\le4$, the reference state contains at most two qubits (see Eq.~\eqref{eqb1}), where any basis permutation is Clifford. 
Since SRE is Clifford invariant, the reference-state SRE is a symmetric function of the Schmidt coefficients and can be expressed through low-order R\'enyi entropies. 
For higher ranks, generic basis permutations are no longer Clifford operations. The simplest example is the Toffoli gate. 
Consequently, the SRE depends on the binary assignment of the ordered coefficients and the associated XOR structure. 
}

\section{Nonlocal SRE in quantum many-body sytems}\label{sec3}
In this section, we investigate the nonlocal magic resource (NL) or nonlocal SRE in quantum many-body systems. Since directly optimizing the nonlocal SRE becomes computationally prohibitive for large entanglement spectra, our analysis for large-rank states is based on the conjecture proposed in Eq.~\eqref{conjecture}, namely that the nonlocal SRE is given by the SRE of the reference state, which can be efficiently evaluated using XOR-FWHT algorithms~\cite{Huang2026,Sierant2026}. For small Schmidt ranks, where numerical optimization remains feasible, we find that the optimized values are consistent with the reference-state SRE. Therefore, throughout this section, we do not distinguish between $M_2^{\mathrm{NL,ref}}$ and $M_2^{\mathrm{NL}}$.

\subsection{Haar random states}

\begin{figure}
    \includegraphics[width=\linewidth]{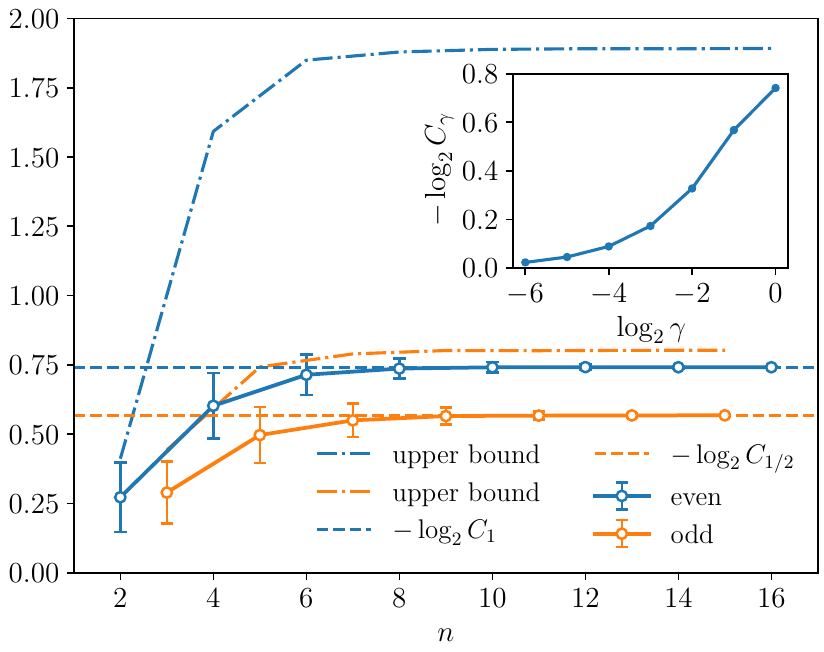}
    \caption{
    Average nonlocal SRE of Haar-random states as a function of system size $n$. 
    For even $n$, we use a balanced bipartition with $\gamma=1$, while for odd $n$, we take $\gamma=1/2$. 
    The inset shows the rapid decay of the average nonlocal SRE away from a balanced bipartition in the thermodynamic limit, as computed from Eq.~\eqref{Cgamma}. 
    For $n\le 10$, the manifold-optimization results agree with our conjecture. 
    The dash-dotted lines indicate the upper bounds in Eq.~\eqref{upperbound}, obtained from Ref.~\cite{Cao2025}, for even (blue) and odd (orange) system sizes. 
    \label{fig3}
    }
\end{figure}

We first examine Haar-random states. As shown in Refs.~\cite{Cao2025,Xiao2026,Gianpaolo2026,Fabio2026}, 
Haar-random states possess only a small amount of nonlocal nonstabilizer resource, 
since their entanglement spectra are close to the flat spectrum. In particular, 
Ref.~\cite{Cao2025} derived the following upper bound for the nonlocal SRE, 
\begin{equation}
    M_2^{\mathrm{NL}}\le\min\{2S_2,4(S_{\max}-S_{1/2})\},\label{upperbound}
\end{equation}
where $S_{\max}=n_A\log2$ and $S_{\alpha}$ denotes the $\alpha$-R\'enyi entanglement entropy. 
Since we use $\log_2$ throughout this work, we have $S_{\max}=n_A$, where $n_A$ is the number of qubits in subsystem $A$, and we assume $n_A\le n_B$.
Based on our conjecture, we further show that the nonlocal SRE of 
bipartite Haar-random states approaches an $O(1)$ constant 
in the thermodynamic limit. More precisely, for $d_A\le d_B$ and fixed aspect 
ratio $\gamma=d_A/d_B=2^{n_A-n_B}$ we obtain 
\begin{equation}
    \lim_{n\to+\infty}\underset{\ket{\psi}\sim \mu_H}{\mathbb{E}}\left[M_2^{\mathrm{NL}}(\ket{\psi}_{AB})\right]=-\log_2C_{\gamma}. 
\end{equation}
Here $\mu_H$ denotes the Haar measure. 
The constant $C_\gamma$ depends only on the ratio of the subsystem Hilbert-space dimensions and is given by the dyadic affine-cube average, 
\begin{equation}
    C_\gamma=\int_{[0,1]^4}\prod_{\omega\in\{0,1\}^3}g_\gamma(u\oplus\omega_1v\oplus\omega_2w\oplus\omega_3z)\mathrm{d}u\mathrm{d}v\mathrm{d}w\mathrm{d}z,
    \label{Cgamma}
\end{equation}
where $\oplus$ denotes dyadic bitwise addition on $[0,1]$. 
See Appendix \ref{app:nonlocal_haar} for detailed derivation. 
The function $g_\gamma(u)$ is defined as the square root of the descending quantile function of 
the Marchenko-Pastur distribution~\cite{Zyczkowski2011}. Explicitly, it satisfies 
\begin{equation}
    \int_{g^2_{\gamma}(u)}^{x_+}\frac{\sqrt{(x_+-x)(x-x_-)}}{2\pi\gamma x}\mathrm{d}x=u, 
\end{equation}
with $x_{\pm}=(1\pm\sqrt{\gamma})^2$. From Eq.~\eqref{Cgamma}, it is obvious that the average nonlocal SRE over bipartite Haar random state converges to a finite value. But the analytical evaluation of $C_\gamma$ appears difficult due to the dyadic affine-cube structure. 
We therefore evaluate $C_\gamma$ numerically for different values of $\gamma$, as shown in the inset of Fig.~\ref{fig3}, which can be regarded as the Page curve for nonlocal SRE~\cite{Page1993,Foong1994,Sen1996}.

Haar-random states provide a useful example for distinguishing 
full-state nonstabilizerness from its genuinely nonlocal 
component. A generic Haar-random state is expected to have a 
large SRE in a fixed computational basis, reflecting the 
highly non-Clifford structure of its full wave function~\cite{Turkeshi202502}. 
However, after minimizing over local unitary transformations, 
the nonlocal SRE is controlled only by the entanglement 
spectrum. Since the Haar entanglement spectrum is close 
to the flat spectrum, with only finite Marchenko-Pastur 
fluctuations, the nonlocal SRE approaches an $O(1)$ constant 
rather than growing extensively with subsystem size. 
This indicates that most of the nonstabilizerness of a 
Haar-random state is removable by local unitary transformations. 
In this sense, Haar-random states are highly magical as 
full many-body wave functions, while their irreducible 
nonlocal magic content remains weak. 

\subsection{Ising model}

\begin{figure*}
    \includegraphics[width=\linewidth]{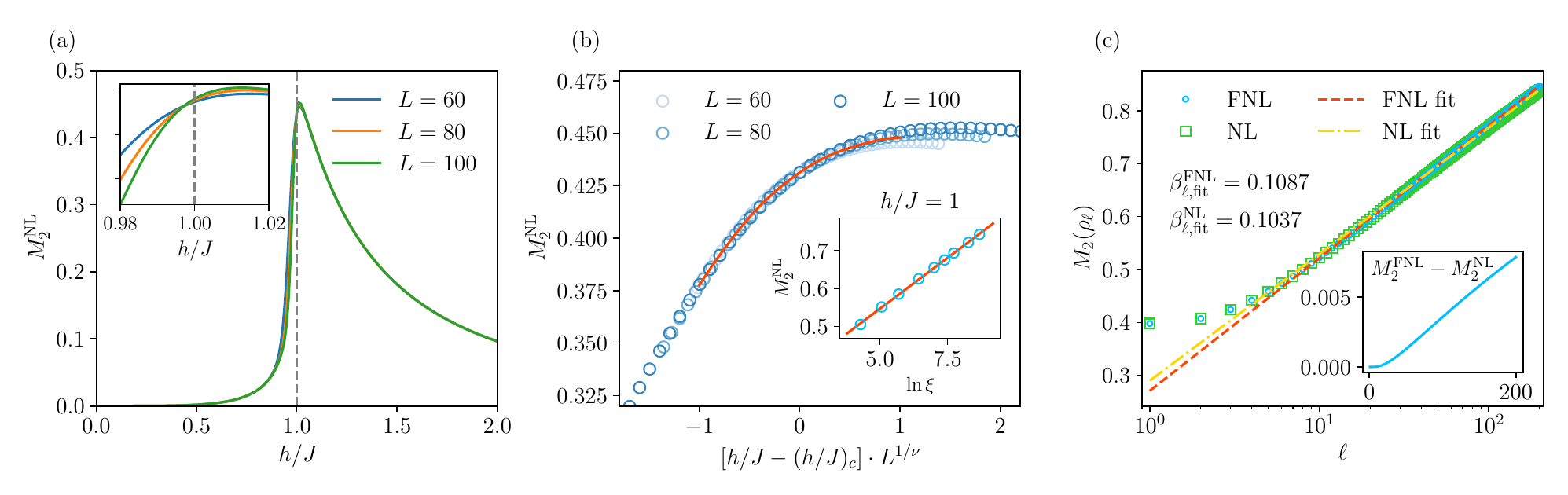}
    \caption{
    Nonlocal SRE in the Ising model.
    For finite systems, we obtain the entanglement spectrum using DMRG, while the thermodynamic-limit results are computed with VUMPS.
    The nonlocal SRE results shown in panels (a) and (b) are independently verified by manifold optimization. 
    (a) Nonlocal SRE of the Ising model under an equal bipartition as a function of $h/J$.
    The inset shows a magnified view near the critical point. 
    (b) Finite-size scaling of the nonlocal SRE. Fixing $\nu=1$, we optimize the position of critical point and obtain $(h/J)_c \approx 0.997$.
    The inset shows the nonlocal SRE at criticality as a function of the correlation length $\xi$ extracted from VUMPS.
    (c) Comparison between the fermionic nonlocal magic resource (FNL) and our nonlocal magic resource (NL) for a subsystem of size $\ell$ in the critical Ising chain.
    The FNL remains larger than the NL for all subsystem sizes considered, as highlighted in the inset, and the difference grows with increasing $\ell$. For $\ell\ge17$, we truncate the entanglement spectrum and only retain Schmidt values larger than $10^{-15}$ to compute NL. 
    \label{fig4}
    }
\end{figure*}

Next, we study the nonlocal SRE in one-dimensional transverse-field Ising model (TFIM), whose Hamiltonian is given by
\begin{equation}
    \hat{H}_{\mathrm{TFIM}}=-h\sum_i \hat{\sigma}^z_i-J\sum_i \hat{\sigma}^x_i \hat{\sigma}^x_{i+1}.
\end{equation}
The TFIM can be exactly solved by mapping it to a free-fermion model through the Jordan-Wigner transformation. 
This allows us to compare our nonlocal SRE with the fermionic nonlocal magic resource (FNL) recently introduced for fermionic Gaussian states~\cite{Mario2026,Daniele2026}. 
We will discuss the relation between FNL and our nonlocal magic resource (NL) in the following.

We first use the density-matrix renormalization group (DMRG) algorithm~\cite{SRWhite1992,itensor} to obtain the ground state of the TFIM under open boundary conditions (OBC). 
As shown in Fig.~\ref{fig4}(a), the nonlocal SRE $M_2^{\mathrm{NL}}$ exhibits a pronounced peak near the critical point. 
Interestingly, for the finite sizes considered, the behavior of $M_2^{\mathrm{NL}}$ resembles that of a Binder cumulant. The curves for different system sizes appear to cross near the phase transition point. 
To examine this observation more quantitatively, we perform a finite-size scaling analysis of the nonlocal SRE, as shown in Fig.~\ref{fig4}(b). 
Fixing the Ising correlation-length exponent to $\nu=1$, we obtain a fitted critical point $(h/J)_c\approx 0.997$, slightly below the exact value $(h/J)_c=1$.

We then use the variational uniform matrix product state (VUMPS) algorithm~\cite{Zauner2018,Vanderstraeten2019} to compute the correlation length $\xi$ and the entanglement spectrum directly in the thermodynamic limit. 
By varying the bond dimension, we extract the scaling behaviour of the nonlocal SRE at the Ising critical point. 
As shown in the inset of Fig.~\ref{fig4}(b), the data are consistent with a logarithmic scaling with the correlation length,
\begin{equation}
    M_2^{\mathrm{NL}}\sim\frac12 \beta_{\xi}^{\mathrm{NL}}\ln \xi ,
    \label{eq20}
\end{equation}
with $\beta_{\xi,\mathrm{fit}}^{\mathrm{NL}}\approx 0.1077$. 
This observation clarifies the apparent Binder-cumulant-like behavior in finite systems. 
Unlike a Binder cumulant, the nonlocal SRE is not a dimensionless quantity whose critical value remains independent of system size. 
If $M_2^{\mathrm{NL}}$ had an exact size-independent crossing at criticality, it would not grow with either the system size or the correlation length. 
Instead, Eq.~\eqref{eq20} indicates that the nonlocal SRE behaves more like the entanglement entropy (EE). It grows logarithmically with the infrared cutoff at criticality. 
Therefore, the crossing observed in finite-size data should be considered as an approximate finite-size feature rather than an asymptotic Binder-cumulant-type scaling.
We note, however, that this conclusion is limited by the accessible range of correlation lengths. 
Since $M_2^{\mathrm{NL}}$ depends on the sorted many-body entanglement spectrum, larger bond dimensions may reveal additional crossover effects associated with the high-lying part of the spectrum. Thus, although the present data are consistent with logarithmic growth, we cannot completely exclude eventual saturation at much larger $\xi$.

We now compare the FNL $M_{\alpha}^{\mathrm{FNL}}$ with our (reference) NL $M_{\alpha}^{\mathrm{NL,ref}}$ in the TFIM at the critical point. 
We first note that FNL provides a general upper bound for NL,
\begin{equation}
    M_{\alpha}^{\mathrm{FNL}}\geq M_{\alpha}^{\mathrm{NL,ref}} . \label{FNL_NL_compare}
\end{equation}
Importantly, this inequality does not rely on our conjecture and holds generally. 
A proof and further discussion are provided in Appendix~\ref{app4}.

Although FNL can be evaluated efficiently from the reduced Majorana covariance matrix, the direct evaluation of NL is much more demanding. 
In particular, computing NL requires the stabilizer entropy of the sorted entanglement spectrum and becomes prohibitively expensive for large subsystem sizes. 
We therefore compare FNL and NL for subsystem sizes $\ell=1,\dots,200$. 
The entanglement spectrum is obtained from the positive eigenvalues of the reduced Majorana covariance matrix, as detailed in Appendix~\ref{app4}. 
The results are shown in Fig.~\ref{fig4}(c). 
We fit the FNL using the logarithmic form
\begin{equation}
    M_2^{\mathrm{FNL}}\sim\beta_{\ell}^{\mathrm{FNL}}\ln \ell ,
\end{equation}
and obtain $\beta_{\ell,\mathrm{fit}}^{\mathrm{FNL}}\approx 0.1087$, in good agreement with Ref.~\cite{Daniele2026}. 
The inset of Fig.~\ref{fig4}(c) shows the difference between FNL and NL. 
As the subsystem size increases, the gap between the two quantities also grows. 

At present, we do not have an analytic asymptotic formula for NL analogous to the closed-form expression for FNL. 
The reason is that FNL depends only on the single-particle covariance spectrum, whereas NL depends on the full many-body entanglement spectrum together with its ordering structure. 
More specifically, after sorting the Schmidt coefficients in descending order, the nonlocal SRE becomes sensitive to the binary labeling and the associated XOR correlations of the ordered spectrum. 
This information is not captured by the single-particle entanglement energies or their spacings alone. 
Nevertheless, FNL gives a useful and well-controlled upper bound on NL at criticality.

Finally, we note that the fitted coefficients satisfy approximately $\beta_{\ell,\mathrm{fit}}^{\mathrm{FNL}}\approx\beta_{\xi,\mathrm{fit}}^{\mathrm{NL}}$. 
This agreement is not accidental. 
In Fig.~\ref{fig4}(c), the subsystem is a finite interval embedded in an infinite critical chain and is connected to the rest of the system through two entanglement cuts. 
By contrast, in the inset of Fig.~\ref{fig4}(b), the subsystem effectively corresponds to a semi-infinite bipartition, which contains only one entanglement cut. 
Thus, if the nonlocal SRE has a boundary-additive logarithmic scaling, the factor of $1/2$ in Eq.~\eqref{eq20} naturally accounts for the difference between the one-cut and two-cut geometries. We emphasize, again, that the precise asymptotic scaling of NL with $\ell$ or $\xi$ remains an open question, especially because the difference between FNL and NL itself increases with subsystem size. 

\subsection{XXZ chain}

\begin{figure}
    \includegraphics[width=\linewidth]{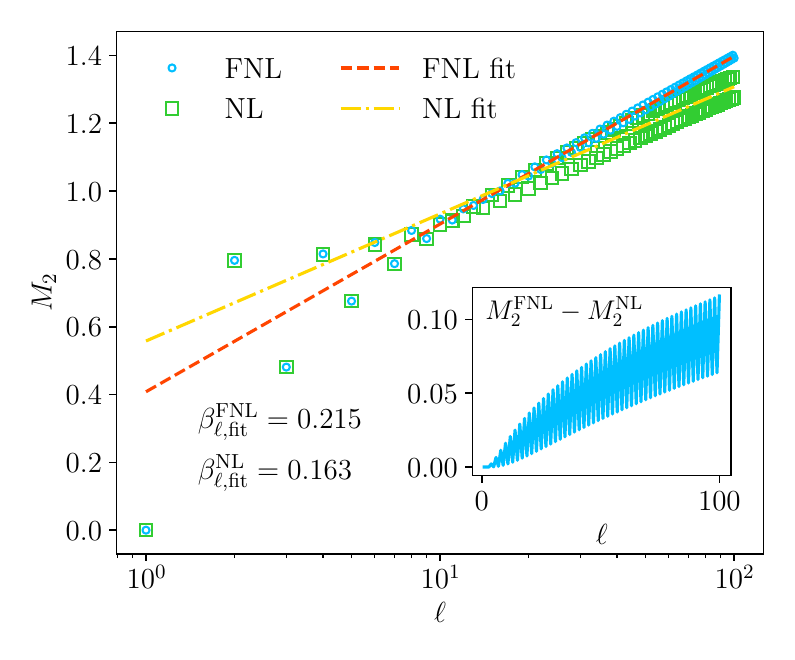}
    \caption{
    FNL and NL as the function of subsustem size $\ell$ in the XXZ chain at $\Delta=0$, corresponding to a free fermion chain. 
    The FNL remains larger than the NL for all subsystem sizes considered, as highlighted in the inset, and the difference grows with increasing $\ell$. For $\ell\ge17$, we truncate the entanglement spectrum and only retain Schmidt values larger than $10^{-15}$ to compute NL. 
    \label{fig5}
    }
\end{figure}

Next, we study the simplest interacting fermion model, the XXZ chain. 
The XXZ-chain Hamiltonian is given by 
\begin{equation}
    \hat{H}_{\mathrm{XXZ}}=
    J\sum_i\left(
    \hat{\sigma}_i^x\hat{\sigma}_{i+1}^x+\hat{\sigma}_i^y\hat{\sigma}_{i+1}^y+
    \Delta\hat{\sigma}_i^z\hat{\sigma}_{i+1}^z\right),
\end{equation}
where $J$ sets the overall energy scale and $\Delta$ denotes the anisotropy. 
Through the Jordan-Wigner transformation, the XXZ chain can be mapped to a one-dimensional spinless fermion model with nearest-neighbor hopping and density-density interactions. 
In this fermionic language, the $\Delta=0$ point corresponds to free fermions, while finite $\Delta$ introduces interactions. 
The ground-state phase diagram of the XXZ chain is well understood. 
For $-1<\Delta\leq 1$, the system is gapless and described by a Luttinger liquid with central charge $c=1$~\cite{Giamarchi2003}. 
For $\Delta>1$, the system enters a gapped antiferromagnetic phase with N\'eel order, while for $\Delta<-1$ the ground state is ferromagnetic and gapped. 
Thus, by tuning $\Delta$, one can continuously move from the free-fermion point to genuinely interacting critical states and further into gapped phases. 
This makes the XXZ chain a useful benchmark for understanding how nonlocal nonstabilizerness behaves in interacting many-body systems.

Unlike the TFIM, whose ground state is a fermionic Gaussian state, the XXZ chain at $\Delta\neq0$ is interacting in the fermionic representation. 
As a result, the fermionic nonlocal magic formula based only on the reduced Majorana covariance matrix is no longer applicable. 
Instead, we compute the entanglement spectrum using tensor-network methods and evaluate the nonlocal SRE from the sorted Schmidt spectrum. 
In the thermodynamic limit, we use VUMPS to obtain the infinite MPS ground state and extract both the correlation length and the entanglement spectrum. 
This allows us to investigate whether the nonlocal SRE exhibits universal scaling behavior in the Luttinger-liquid phase and how it changes across the interacting XXZ phase diagram. 

We begin from the free-fermion case $\Delta=0$. 
At this point, the XXZ chain reduces to the XX model, which can be mapped to a free spinless-fermion hopping model by the Jordan-Wigner transformation. 
Since the ground state is Gaussian, the reduced density matrix of a subsystem is completely determined by its two-point correlation matrix. 
Therefore, for a subsystem of length $\ell$, the entanglement spectrum can be obtained analytically. 

For an infinite XX chain at half filling, the correlation matrix restricted to a contiguous subsystem $A$ is
\begin{equation}
    C_{mn}=
    \langle \hat{c}_m^\dagger \hat{c}_n\rangle=
    \begin{cases}
        \dfrac{\sin[k_F(m-n)]}{\pi(m-n)}, & m\neq n,\\[6pt]
        \dfrac{k_F}{\pi}, & m=n,
    \end{cases}
\end{equation}
with Fermi momentum $k_F=\pi/2$. 
Diagonalizing this $\ell\times \ell$ correlation matrix gives the single-particle entanglement eigenvalues $\{\nu_i\}_{i=1}^{\ell}$. 
The full many-body entanglement spectrum is then generated as
\begin{equation}
    \lambda_{\mathbf n}
    =\prod_{j=1}^{\ell}
    \nu_j^{n_j}
    (1-\nu_j)^{1-n_j},
    \quad
    \mathbf n=(n_1,\dots,n_\ell)\in\{0,1\}^{\ell}.
\end{equation}
Using this exact entanglement spectrum, we calculate the nonlocal SRE $M_2^{\mathrm{NL}}$ from the sorted Schmidt spectrum. 
At the same time, since the state is Gaussian, we can also compute the FNL $M_2^{\mathrm{FNL}}$,
\begin{equation}
    M_2^{\mathrm{FNL}}
    =
    \sum_{j=1}^{\ell}
    m_2(a_j^2), 
\end{equation}
with $a_j^2=(1-2\nu_j)^2$ and $m_2(x)=-\log_2(1-x+x^2)$. 
This allows a direct comparison between the two quantities as a function of the subsystem size $\ell$. 

As shown in Fig.~\ref{fig5}, both NL and FNL increase with $\ell$, where FNL still remains larger than 
NL for all subsystem sizes considered, consistent with the general inequality in Eq.~\eqref{FNL_NL_compare}. 
We further fit the subsystem-size dependence using the logarithmic form
\begin{equation}
    M_2(\ell)\sim\beta_{\ell}\ln \ell,
\end{equation}
for both FNL and NL. 
We note that the FNL coefficient is approximately twice that obtained at the Ising critical point, 
\begin{equation}
    \beta_{\ell,\mathrm{XX}}^{\mathrm{FNL}}\simeq2\beta_{\ell,\mathrm{Ising}}^{\mathrm{FNL}}. 
\end{equation}
This is because the Ising critical point is described by a single massless Majorana fermion, 
whereas the XX chain at half filling is described by a massless Dirac fermion, 
or equivalently two independent massless Majorana fermions. 
In the scaling limit, the reduced covariance matrix of the XX chain decomposes into two Majorana covariance sectors. 
Since the FNL of a fermionic Gaussian state is additive over the single-particle entanglement modes, 
the two Majorana sectors contribute two identical leading logarithmic terms. 
This provides a useful consistency check for the free-fermion calculation and shows that, 
for Gaussian critical systems, the leading FNL coefficient is additive over independent gapless Majorana degrees of freedom, 
which is similar to central charge for EE.

By contrast, the same factor-of-2 argument does not directly apply to $M_2^{\mathrm{NL}}$. 
Although NL is computed from the same many-body entanglement spectrum, it is sensitive to the sorted binary labeling of the Schmidt coefficients and to the associated XOR correlations entering the stabilizer entropy. 
Therefore, its logarithmic coefficient is not determined solely by the number of gapless Majorana modes or by the central charge. 
This is precisely why the comparison between FNL and NL at the free-fermion point is useful where FNL captures the Gaussian-mode contribution in a closed form, while NL probes the nonlocal magic remaining after optimizing the ordering structure of the entanglement spectrum. 

\begin{figure}
    \includegraphics[width=\linewidth]{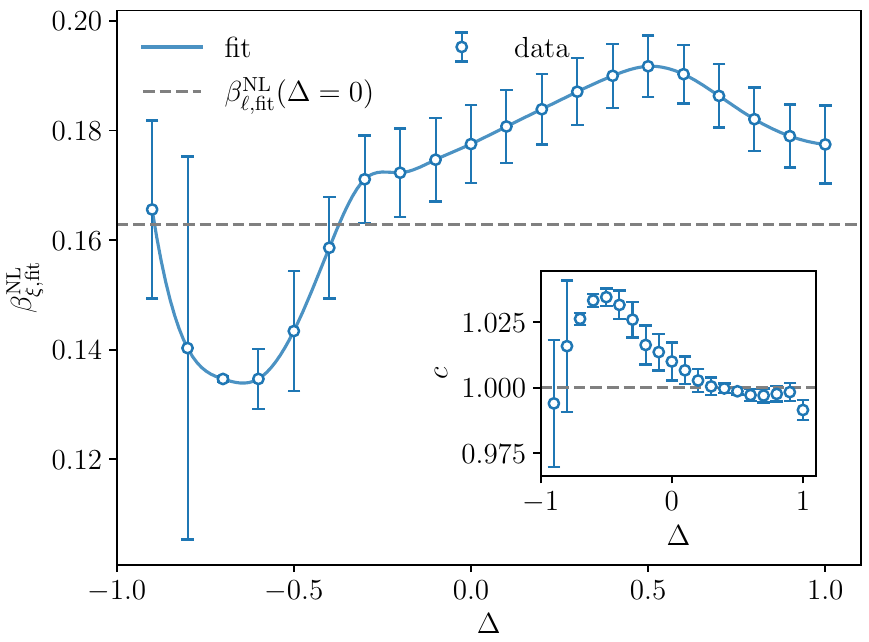}
    \caption{
    Scaling coefficient of the nonlocal SRE in the critical XXZ chain. 
    For each anisotropy $\Delta\in[-0.9,1.0]$, we compute the ground state using VUMPS and fit 
    $\beta_{\xi,\mathrm{fit}}^{\mathrm{NL}}$ according to Eq.~\eqref{xxz_betafit}. 
    The blue circles show the fitted coefficients, and the solid line is a guide to the eye.
    The gray dashed line marks the coefficient $\beta_{\ell,\mathrm{fit}}^{\mathrm{NL}}$ 
    extracted from subsystem-size scaling at the free-fermion point $\Delta=0$.
    The inset shows the central charge extracted from the scaling relation $S_1\sim (c/6)\ln \xi$ with $\xi$ the correlation length.
    \label{fig8}
    }
\end{figure}

We further investigate the dependence of the nonlocal SRE on the XXZ anisotropy $\Delta$ throughout the critical Luttinger-liquid regime. 
For each $\Delta$ from $-0.9$ to $1.0$, we compute the ground state using VUMPS varying the bond dimension from 128 to 1024, 
and fit the scaling of the nonlocal SRE with the EE using 
\begin{equation}
    M_2^{\mathrm{NL}}\sim \frac{1}{2}\beta^{\mathrm{NL}}_{\xi}\ln\xi\sim\frac{3\beta^{\mathrm{NL}}_{\xi}}{c}S_1, \label{xxz_betafit}
\end{equation}
where $S_1$ is the von Neumann entropy and it scales as $(c/6)\ln\xi$ in the critical region with $c$ the central charge~\cite{Giamarchi2003,Vidal200306,Affleck1991,Christoph1994,Pasquale2004,Pasquale2009}. 
As shown in Fig.~\ref{fig8}, the fitted coefficient $\beta_{\xi,\mathrm{fit}}^{\mathrm{NL}}$ varies with $\Delta$. 
This behavior is qualitatively different from that of the EE, whose leading logarithmic coefficient is fixed by the central charge and therefore remains constant throughout the $c=1$ Luttinger-liquid phase. 
In particular, even at the free-fermion point $\Delta=0$, the coefficient of $M_2^{\mathrm{NL}}$ is not simply twice the Ising critical value, in contrast to the fermionic nonlocal magic resource, whose leading coefficient is additive over independent Majorana modes. 
For interacting points $\Delta\neq0$, the coefficient changes further with the interaction strength, although the central charge remains unchanged. 
This demonstrates that nonlocal SRE is sensitive to finer structures of the entanglement spectrum than conventional entanglement entropies and FNL. 
The reason is that, after sorting the many-body Schmidt coefficients, $M_2^{\mathrm{NL}}$ depends not only on their distribution but also on their binary assignment and the associated XOR correlations entering the stabilizer entropy. 
Thus, the variation of $\beta_{\xi}^{\mathrm{NL}}$ across the XXZ critical phase suggests that nonlocal nonstabilizerness encodes interaction-dependent information in the entanglement spectrum beyond the central charge.

\subsection{PXP model}

Finally, we consider a more experimentally accessible model, the PXP model realized in Rydberg atom arrays~\cite{Choi2019,Bluvstein2021,Fendley2004,Lesanovsky2012,Turner2018,Khemani2019}. 
The PXP Hamiltonian is given by
\begin{equation}
    \hat{H}_{\mathrm{PXP}}=\sum_i\hat{P}_{i-1}\hat{X}_i\hat{P}_{i+1},
\end{equation}
where $\hat{P}_i=\ket{0}_i\bra{0}$ is the projector onto the empty state and 
$\hat{X}_i=\ket{0}_i\bra{1}+\ket{1}_i\bra{0}$ is the Pauli-$x$ operator. 
Here $\ket{0}$ and $\ket{1}$ denote the empty and occupied Rydberg states, respectively. 
The projectors in the Hamiltonian encode the Rydberg blockade constraint. 
The atom can be flipped only when its two neighboring sites are both in the empty state.

The PXP model has been extensively studied as a paradigmatic example of quantum many-body scars and weak ergodicity breaking~\cite{Choi2019}. 
Starting from generic product states, such as the empty state $\ket{00\cdots}$, the dynamics are expected to be rapidly thermalizing, resembling those of a chaotic many-body system. 
In contrast, for certain simple initial states, most notably the period-2 state $\ket{1010\cdots}$ and the period-3 state $\ket{100100\cdots}$, the system exhibits anomalously slow thermalization~\cite{Turner2018oct}. 
These special initial states show long-lived revivals in the Loschmidt echo and a slow, approximately logarithmic growth of the EE. 
This non-thermalizing behavior originates from a set of atypical, weakly entangled eigenstates embedded in the many-body spectrum. 
These eigenstates are approximately equally spaced in energy and have anomalously large overlap with the special product states, leading to coherent oscillations in the quench dynamics. 
They are known as quantum many-body scar states~\cite{Turner2018,Chandran2023,Serbyn2021}. 
See also Appendix~\ref{app:pxp} for related discussion. 

\begin{figure}
    \includegraphics[width=\linewidth]{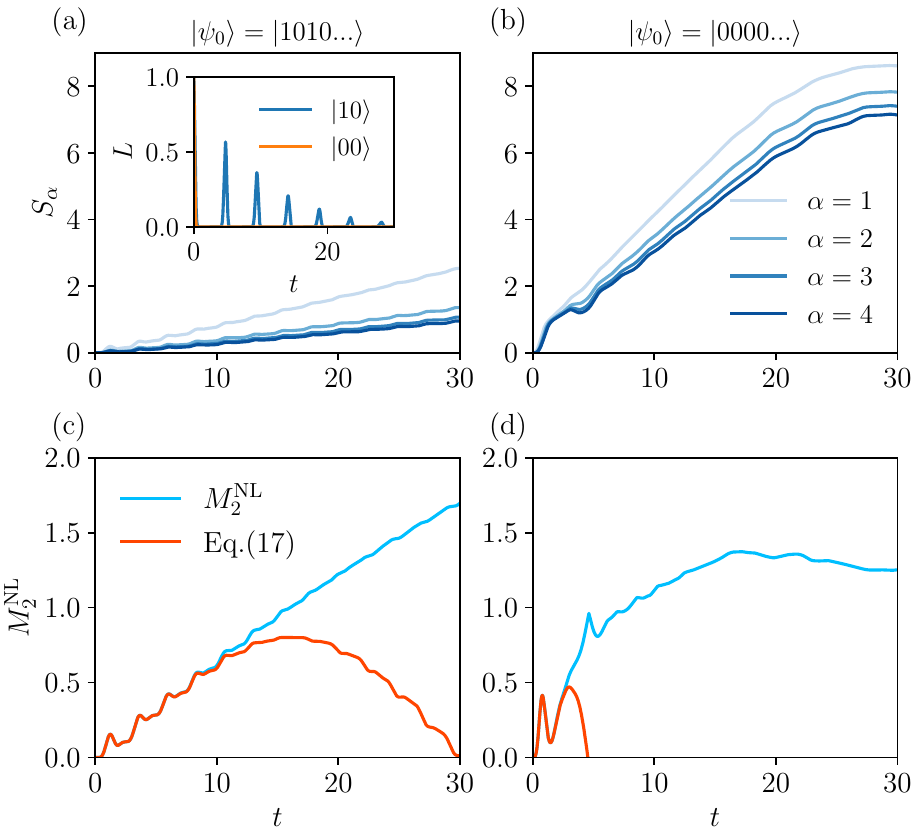}
    \caption{
    Entanglement and nonlocal magic dynamics in the PXP model with system size $N=30$ under OBC.
    (a,b) Time evolution of the R\'enyi entanglement entropies $S_\alpha$ with $\alpha=1,2,3,4$ for the initial states $\ket{\psi_0}=\ket{1010\cdots}$ and $\ket{\psi_0}=\ket{0000\cdots}$, respectively.
    The inset in (a) shows the Loschmidt echo $L(t)=|\braket{\psi_0|e^{-iHt}|\psi_0}|^2$.
    (c,d) Time evolution of the nonlocal SRE $M_2^{\mathrm{NL}}$ for the same two initial states.
    The blue curve denotes the accurate nonlocal SRE, while the orange curve denotes the low-rank approximated nonlocal SRE obtained from Eq.~\eqref{lowrankapprox}, which only requires the R\'enyi entropies $S_2$, $S_3$, and $S_4$.
    \label{fig6}
    }
\end{figure}

We first study the short-time behavior of the PXP model. 
From the perspective of EE, the scarred and non-scarred initial states exhibit clearly different dynamics. 
As shown in Fig.~\ref{fig6}(a), for the scarred initial state $\ket{1010\cdots}$, the R\'enyi entanglement entropies $S_\alpha$ grow slowly and remain relatively small within the time window considered. 
The inset shows pronounced revivals in the Loschmidt echo, confirming the well-known scarred dynamics of the PXP model. 
By contrast, Fig.~\ref{fig6}(b) shows that the empty initial state $\ket{0000\cdots}$ generates entanglement much more rapidly, and the R\'enyi entropies reach substantially larger values. 
Thus, at the level of conventional EE, the scarred initial state appears much less entangling than the empty initial state.

The nonlocal SRE, however, reveals a rather different aspect of the dynamics. 
As shown in Fig.~\ref{fig6}(c), the scarred initial state $\ket{1010\cdots}$ exhibits a rapid growth of $M_2^{\mathrm{NL}}$. 
In particular, at short times, $M_2^{\mathrm{NL}}$ grows approximately linearly in time. 
This behavior is in sharp contrast to the slow growth of the entanglement entropies shown in Fig.~\ref{fig6}(a). 
Therefore, although the scarred trajectory generates only a relatively small amount of bipartite entanglement, its entanglement carries a substantial amount of nonlocal magic resource. 
This suggests that quantum many-body scars do not simply suppress all forms of quantum complexity. 
Instead, scarred dynamics can efficiently generate nonstabilizer correlations within a restricted, weakly entangled dynamical subspace.
For the empty initial state $\ket{0000\cdots}$, the behavior is different. 
As shown in Fig.~\ref{fig6}(d), the nonlocal SRE grows faster than in the scarred case but reaches a saturated or quasi-saturated regime at relatively early times. 
Compared with Fig.~\ref{fig6}(b), we can conclude that rapid entanglement growth does not necessarily imply rapid growth of nonlocal magic resource. 
The empty state produces large bipartite entanglement, but the corresponding nonlocal SRE remains comparatively modest and saturates earlier. 
This separation between entanglement growth and nonlocal magic resource growth highlights that $M_2^{\mathrm{NL}}$ captures information beyond the amount of entanglement encoded in the Schmidt spectrum.

We also compare the full-rank nonlocal SRE with the low-rank approximated nonlocal SRE as defined in Eq.~\eqref{lowrankapprox}, shown by the orange curve in Fig.~\ref{fig6}(c) and (d). 
This approximation is experimentally motivated. Instead of reconstructing the full entanglement spectrum, it only requires the low-order R\'enyi entropies $S_2$, $S_3$, and $S_4$, which can be measured in experiments~\cite{Daley2012,Abanin2012,Islam2015,Zhou2026}.
For the scarred initial state, the low-rank approximation captures the early-time growth of $M_2^{\mathrm{NL}}$ reasonably well, indicating that at short times, the entanglement spectrum is still effectively dominated by a few leading Schmidt values~\cite{Smith2025,Michailidis2020,Ho2019}. 
At later times, the approximation deviates from the full-rank nonlocal SRE, implying that higher-rank components of the entanglement spectrum and their ordering structure become increasingly important. 
Thus, the low-rank approximation provides an experimentally accessible probe of nonlocal magic resource in the early-time regime, while the full nonlocal SRE remains sensitive to the detailed many-body structure of the entanglement spectrum.

\begin{figure}
    \includegraphics[width=\linewidth]{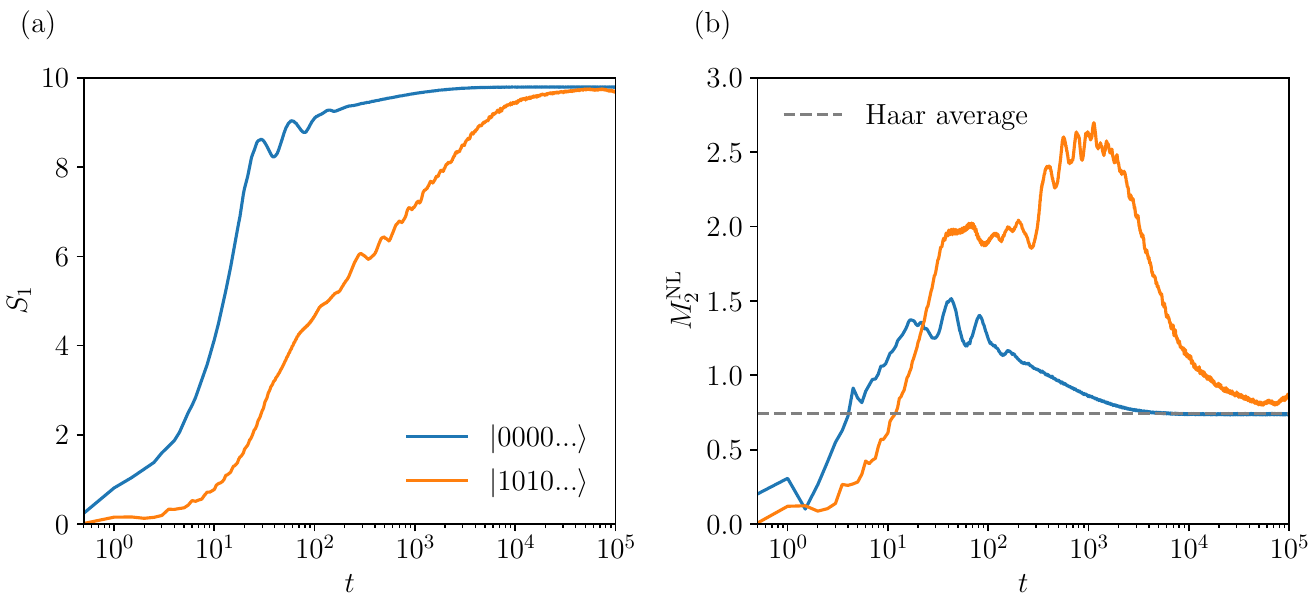}
    \caption{
    Long-time dynamics of the PXP model starting from the initial states $\ket{0000\cdots}$ and $\ket{1010\cdots}$ for a system of size $N=30$ with OBC. 
    We show the entanglement entropy $S_1(t)$ in (a) and the nonlocal SRE in (b).
    \label{fig7}
    }
\end{figure}

We then study the long-time dynamics of the PXP model starting from these two different initial product states. 
As shown in Fig.~\ref{fig7}(a), the empty initial state $\ket{0000\cdots}$ exhibits a rapid growth of the EE. 
The entropy grows approximately linearly at early and intermediate times and quickly approaches a saturated value. 
By contrast, the scarred initial state $\ket{1010\cdots}$ shows much slower entanglement growth. 
Over a broad time window, the EE grows only logarithmically in time, reflecting the slow thermalization associated with quantum many-body scar dynamics. 

However, the behavior of the nonlocal SRE is markedly different. 
For the empty initial state, despite the rapid growth and early saturation of the EE, the nonlocal SRE grows much more slowly. 
Its early-time growth is closer to logarithmic rather than linear, indicating that fast entanglement production does not necessarily generate irreducible nonstabilizerness efficiently. 
After reaching a maximum, $M_2^{\mathrm{NL}}$ decreases slowly and appears to approach the Haar-random value at long times. 
This suggests that, although the empty state rapidly explores highly entangled states, the corresponding entanglement spectrum gradually becomes closer to that of a typical random state, for which the nonlocal SRE remains relatively small.
In contrast, the scarred initial state displays the opposite trend. 
Although its EE grows slowly, the nonlocal SRE increases rapidly and is approximately linear at early times. 
After this initial growth, $M_2^{\mathrm{NL}}$ develops a plateau, indicating that the scarred dynamics first generates a relatively stable amount of irreducible nonstabilizerness within a weakly entangled subspace. 
At later times, as the EE continues to grow, the nonlocal SRE shows a step-like increase from a lower plateau to a higher plateau. 
This behavior suggests that the growth of nonlocal SRE is sensitive not only to the amount of entanglement, but also to qualitative rearrangements of the entanglement spectrum. 
Eventually, when the EE approaches its saturated regime, the nonlocal SRE also starts to decrease slowly and tends toward the Haar-random value. 

These results demonstrate a strong separation between entanglement growth and nonlocal nonstabilizerness growth in constrained many-body dynamics. 
Starting from the empty initial state, entanglement is generated rapidly, whereas the nonlocal SRE grows slowly and remains relatively modest. 
By contrast, the scarred initial state exhibits slow entanglement growth but efficiently generates a larger and longer-lived nonlocal SRE, with pronounced plateau and step-like structures at long times. 
An eigenstate-resolved analysis further shows that this enhancement does not originate from unusually large nonlocal SRE of the scarred eigenstates themselves, whose values remain comparable to the Haar-random average. 
Instead, it is associated with the coherent superposition and long-lived interference of the scarred eigenstates during the nonequilibrium evolution. 
See Appendix~\ref{app:pxp} for further discussion. 
Therefore, nonlocal SRE captures dynamical information beyond conventional entanglement entropy, probing not only the amount of bipartite entanglement but also the detailed structure and ordering of the entanglement spectrum.

\section{Discussion and outlook}\label{sec4}

In this work, we proposed a Schmidt-reference-state framework that evaluates nonlocal nonstabilizerness directly from the sorted entanglement spectrum, bypassing the highly nonconvex local-unitary optimization. 
The conjectured formula is supported by the first-order optimality condition of the reference state and by manifold-optimization benchmarks for rank-4 spectra and representative many-body states.
The main physical message is that nonlocal SRE is constrained by the Schmidt spectrum, but is not a simple function of entanglement entropy or Gaussian covariance data for free fermions. 
Haar-random states have large full-state nonstabilizerness but only small nonlocal SRE, reflecting the near-flatness of their Schmidt spectra. In critical spin chains, nonlocal SRE shows logarithmic behavior, yet its coefficient is sensitive to structures beyond the central charge. 
In the XXZ chain, this coefficient varies across the critical region even though $c=1$ remains fixed, and at the free-fermion point it does not follow the factor-of-2 relation obeyed by FNL. 
This indicates that NL probes the ordered many-body entanglement spectrum, including its binary assignment and XOR correlations.
Out of equilibrium, the PXP model reveals a clear separation between entanglement growth and nonlocal nonstabilizerness growth. 
The empty initial state exhibits approximately linear entanglement growth but generates only a small amount of nonlocal SRE. 
In contrast, the scarred initial state shows logarithmic entanglement growth while producing a larger and longer-lived nonlocal SRE. 
Therefore, nonlocal SRE provides a complementary diagnostic of constrained dynamics and weak ergodicity breaking.

Several open directions remain. 
The most important theoretical challenge is to prove the Schmidt-reference-state conjecture entirely. 
A complete proof would require understanding the global SRE landscape under local unitary transformations, possibly through Hessian analysis, Pauli-spectrum algebra, or polynomial optimization over local-unitary invariants. 
Such a result would promote the conjectured spectral expression to an exact formula for nonlocal SRE.
It is also important to develop a scaling theory of nonlocal SRE in critical many-body systems, as what has been done for full state SRE~\cite{Hoshino2026prl,Hoshino2026prx,Ryota2026}. 
Our Ising and XXZ results indicate that nonlocal SRE can grow logarithmically, but its coefficient is not determined solely by the central charge. 
In the XXZ chain, this coefficient varies across the Luttinger-liquid phase even though $c=1$ remains fixed, suggesting that nonlocal SRE probes finer interaction-dependent structures of the entanglement spectrum. 
Understanding whether this coefficient is related to the Luttinger parameter or other conformal data is a promising direction.

Our framework is naturally suited for tensor-network calculations. 
Since matrix-product states directly provide the Schmidt spectrum~\cite{Vidal2003}, nonlocal SRE can be evaluated efficiently in one-dimensional weakly entangled systems. 
Extensions to projected entangled-pair states and higher-dimensional tensor networks may enable studies of nonlocal nonstabilizerness in frustrated systems~\cite{Diep2005}, symmetry-protected topological phases~\cite{Wen2012,Senthil2015}, and topologically ordered phases~\cite{Broholm2020,Savary2016,Wen2013,Kitaev2006}. 
Out of equilibrium, nonlocal SRE can be used to diagnose the spreading of irreducible nonstabilizerness in many-body localized systems~\cite{Abanin2019,Falcao2025,Zhang2026}, quench dynamics~\cite{Tirrito2025,Mitra2018}, and random quantum circuits~\cite{Turkeshi2025,Maity2026,Zhang2026}.
From an experimental perspective, the low-rank approximation suggests a practical route to probing nonlocal nonstabilizerness without reconstructing the full entanglement spectrum. 
When the Schmidt spectrum is dominated by a few leading values, the nonlocal SRE can be estimated from low-order R\'enyi entropies. 
This may be useful for programmable quantum simulators such as Rydberg atom arrays~\cite{Bluvstein2021}, trapped ions~\cite{Linke2018}, and superconducting qubits~\cite{Tiff2019}. 
These directions may establish nonlocal SRE as a practical diagnostic of irreducible nonstabilizer correlations in quantum many-body matter.

\begin{acknowledgments}
We thank the useful discussion with ChunJun Cao and Gong Cheng. 
This work is supported by the National Key Research and Development 
of China (Grant No. 2021YFA1402001) and the National Natural Science Foundation of China (NSFC) (Grant No. 12375007). 
\end{acknowledgments}

\section*{Data availability}
The data that support the findings of this article are not publicly available. 
The data are available from the authors upon reasonable request. 

\appendix

\section{Proof of Eq.~\eqref{eq8} and further discussion}\label{app1}
For the entanglement spectrum of a general quantum many-body states $\{\lambda_i\}_{i=0}^{\chi-1}$ 
with $\sum_i\lambda_i=1$ and $\lambda_0\ge\lambda_1\ge ...\ge \lambda_{\chi-1}$, we construct a 
reference state 
\begin{equation}
    \ket{\phi}_{AB}=\sum_{i=0}^{\chi-1}\sqrt{\lambda_i}\ket{c_i}_{A}\ket{c_i}_B. \label{eqa1}
\end{equation}
This is the same as Eq.~\ref{eq4}. We just use a different notation for simplicity. 

Define $H = H_A\otimes \mathbb{I}_B+\mathbb{I}_A\otimes H_B$ as a Hermitian matrix, then the corresponding unitary 
generated by $H$ is 
\begin{equation}
    U_{AB}(t)=e^{-iHt} =U_A(t)\otimes U_B(t)=e^{-iH_At}\otimes e^{-iH_Bt}.
\end{equation}
The $\alpha$-SRE of a general quantum state is 
\begin{equation}
    M_{\alpha}(\ket{\psi})=\frac{1}{1-\alpha}\Big(\log F_{\alpha}(\ket{\psi})-n\log 2\Big),
\end{equation}
where $F_{\alpha}(\ket{\psi})$ is 
\begin{equation}
    F_{\alpha}(\ket{\psi}) = \sum_{P\in\mathcal{P}_n}\braket{\psi|P|\psi}^{2\alpha}. 
\end{equation}
To prove Eq.~\eqref{eq8}, we only need to prove 
\begin{equation}
    \frac{\partial F_{\alpha}(t)}{\partial t}\Bigg|_{t=0}=0, \label{eqa5}
\end{equation}
where we use $F_{\alpha}(t)=F_{\alpha}(U_{AB}(t)\ket{\phi}_{AB})$ for simplicity. 

Now we show Eq.~\eqref{eqa5} is indeed true. Direct calculations yield 
\begin{equation}
    \frac{\partial F_{\alpha}(t)}{\partial t}\Bigg|_{t=0} = 2i\alpha\braket{\phi|[H_{AB}, G]|\phi}
\end{equation}
where $p_P=\braket{\phi|P|\phi}$ and 
\begin{equation}
    G = \sum_{P\in\mathcal{P}_n}p_P^{2\alpha-1}P. 
\end{equation}
Denoting $\rho=\ket{\phi}\bra{\phi}$, we have 
\begin{equation}
    \braket{\phi|[H_{AB}, G]|\phi}=\operatorname{tr}(\rho[H_A,G])+\operatorname{tr}(\rho[H_B,G]). \label{eqa8}
\end{equation}
Take the first term of Eq.~\eqref{eqa8} as an example, we have
\begin{equation}
    \operatorname{tr}(\rho[H_A,G])=\operatorname{tr}(H_A[G,\rho])=\operatorname{tr}_A[H_A\operatorname{tr}_B([G,\rho])].
\end{equation}
Let's carefully examine $\operatorname{tr}_B([G,\rho])$. The reference state is 
\begin{equation}
    \rho=\ket{\phi}\bra{\phi}=\sum_{i,j=0}^{\chi-1}\sqrt{\lambda_i\lambda_j}\ket{c_ic_i}\bra{c_jc_j}, 
\end{equation}
and by definition $G$ only contains the Pauli strings that map $\operatorname{span}\{\ket{c_ic_i}\}$ to $\operatorname{span}\{\ket{c_ic_i}\}$, so inside this subspace $\mathcal{H}_{\mathrm{pair}}=\operatorname{span}\{\ket{c_ic_i}\}$, the $G$ operator can be written as 
\begin{equation}
    G = \sum_{i,j=0}^{\chi-1}g_{ij}\ket{c_ic_i}\bra{c_jc_j},
\end{equation}
so now we have 
\begin{equation}
    \operatorname{tr}_B([G,\rho])=\sum_{i,j=0}^{\chi-1}\sqrt{\lambda_i\lambda_j}(g_{ij}-g_{ji})\ket{c_i}_A\bra{c_i}. 
\end{equation}
The matrix $G$ is hermitian, which gives $g_{ij}=g_{ji}^*$. If $G$ contains imaginary number, then it must come from 
Pauli strings with odd number of $\sigma_y$, and for this kind of $P$, the elements of it are all imaginary, so 
\begin{equation}
    \braket{\phi|P|\phi}=\sum_{i,j=0}^{\chi-1}\sqrt{\lambda_i\lambda_j} \braket{c_ic_i|P|c_jc_j}
\end{equation}
must be imaginary. But $P$ is hermitian, so $\braket{\phi|P|\phi}$ must be real and consequantly $\braket{\phi|P|\phi}$ 
can only be 0 for imaginary $P$. This is why $G$ is real, which gives $g_{ij}=g_{ji}$ so $\operatorname{tr}_B([G,\rho])=0$ and $\operatorname{tr}(\rho[H_A,G])=0$. 
Similarly, we obtain $\operatorname{tr}(\rho[H_B,G])=0$, so $\braket{\phi|[H_{AB}, G]|\phi}=0$ and therefore 
Eq.~\eqref{eqa5} is true. 

We note that in this proof we don't use the condition that the Schmidt values are in descending order. 
So any state that has the form of Eq.~\eqref{eqa1} or can be transformed to Eq.~\eqref{eqa1} via local Cliffords 
satisfty Eq.~\eqref{eq8} and Eq.~\eqref{eqa5}. 
One can show that among these states, the one with descending order of Schmidt values reaches the minimum value 
of $\alpha$-SRE, as shown in Ref.~\cite{Cao2025}. 

We also attempt to prove the Hessian matrix of the reference state is semi-negative with respect to $F_{\alpha}$. 
However, we find it is difficult to prove. Here we give the expression of the Hessian matrix. 
Denoting 
\begin{equation}
    \boldsymbol{\mathrm{H}}(H,G)=\frac{\partial^2F_{\alpha}(t,s)}{\partial s\partial t} \Bigg|_{t=s=0}, 
\end{equation}
where 
\begin{equation}
    F_{\alpha}(t,s)=F_{\alpha}\left(
    e^{-iH_At}e^{-iG_A s}\otimes e^{-iH_Bt}e^{-iG_Bs}\ket{\phi}\right), 
\end{equation}
and 
\begin{eqnarray}
    H = H_A\otimes\mathbb{I}_B+\mathbb{I}_A\otimes H_B,\\
    G = G_A\otimes\mathbb{I}_B+\mathbb{I}_A\otimes G_B,
\end{eqnarray}
we obtain 
\begin{eqnarray}
    \boldsymbol{\mathrm{H}}(H,G)&=
    -2\alpha\displaystyle\sum_{P\in\mathcal{P}_n}p_P^{2\alpha-2}\big[(2\alpha-1)\braket{[H,P]}\braket{[G,P]}\nonumber\\
    &+p_P\braket{[G,[H,P]]}\big]. \label{eqa18}
\end{eqnarray}
Here $\braket{\cdot}$ denotes $\braket{\phi|\cdot|\phi}$. 
It is difficult to analytically prove that $\boldsymbol{\mathrm{H}}(H,G)\le0$ for any $H,G$. Numerically, we test 
a few thousands random $H$ and $G$ for reference states in small systems with automatic differentiation, and do not find $\boldsymbol{\mathrm{H}}(H,G)>0$ up to numerical precision.

\section{Bound on the SRE error of truncated spectrum}\label{app:truncation_error}
In this section, we define the reference state as 
\begin{equation}
    \ket{\tilde{\psi}}=\sum_i\sqrt{\lambda_i}\ket{c_i}, \label{eqb1}
\end{equation}
This state has the same SRE as the one we defined in Eq.~\eqref{eq4}, 
because the two are related by Clifford operations and stabilizer ancillas. 
Explicitly, by appending $n$ stabilizer qubits and applying CNOT gates from the first register to the second, one maps
\begin{equation}
    \sum_i\sqrt{\lambda_i}\ket{c_i}_A\ket{0}_B
    \longmapsto
    \sum_i\sqrt{\lambda_i}\ket{c_i}_A\ket{c_i}_B.
\end{equation}
Since SRE is invariant under Clifford unitaries and unchanged by tensoring stabilizer states, both states have identical SRE.

Let $K$ be the set of retained Schmidt coefficients and denote $D$
the discarded set, with total discarded weight
$\eta=\sum_{i\in D}\lambda_i$. 
The normalized truncated state is
\begin{equation}
    \ket{\tilde{\psi}_{\mathrm{trunc}}}=
    \frac{1}{\sqrt{1-\eta}}
    \sum_{i\in K}\sqrt{\lambda_i}\ket{c_i}.
\end{equation}
Then the overlap between the exact and truncated reference states is 
\begin{equation}
    \braket{\tilde{\psi}|\tilde{\psi}_{\mathrm{trunc}}}
    =\frac{1}{\sqrt{1-\eta}}
    \sum_{i\in K}\lambda_i=
    \sqrt{1-\eta}.
\end{equation}
Therefore, for the corresponding pure-state density matrices
\begin{equation}
    \rho=\ket{\tilde{\psi}}\bra{\tilde{\psi}},
    \quad
    \rho_{\mathrm{trunc}}
    =
    \ket{\tilde{\psi}_{\mathrm{trunc}}}
    \bra{\tilde{\psi}_{\mathrm{trunc}}},
\end{equation}
the trace distance is
\begin{equation}
    T(\rho,\rho_{\mathrm{trunc}})
    =
    \frac12\|\rho-\rho_{\mathrm{trunc}}\|_1
    =
    \sqrt{
    1-
    \left|
    \braket{\tilde{\psi}|\tilde{\psi}_{\mathrm{trunc}}}
    \right|^2
    }
    =
    \sqrt{\eta}.
\end{equation}

We now bound the change of the stabilizer purity $\zeta_{\alpha}$, 
which is defined as 
\begin{equation}
    \zeta_\alpha(\ket{\psi})
    =\frac{1}{d}\sum_{P\in\mathcal{P}_d}\braket{\psi|P|\psi}^{2\alpha}.
\end{equation}
Since $P$ is unitary and Hermitian, one has $|\braket{\psi|P|\psi}|\le1$. 
Denoting
\begin{equation}
    p_P=\braket{\tilde{\psi}|P|\tilde{\psi}},\quad 
    p_P'=\braket{\tilde{\psi}_{\mathrm{trunc}}|P|\tilde{\psi}_{\mathrm{trunc}}}, 
\end{equation}
we consider the difference between $\zeta_{\alpha}(\ket{\tilde{\psi}})$ and $\zeta_{\alpha}(\ket{\tilde{\psi}_{\mathrm{trunc}}})$. 
Using
\begin{equation}
    x^{2\alpha}-y^{2\alpha}=(x-y)
    \sum_{k=0}^{2\alpha-1}x^{2\alpha-1-k}y^k ,
\end{equation}
we obtain 
\begin{align}
    &\left|
    \zeta_\alpha(\ket{\tilde{\psi}})
    -\zeta_\alpha(\ket{\tilde{\psi}_{\mathrm{trunc}}})
    \right|
    \nonumber\\
    &\le
    \frac{1}{d}
    \sum_{P\in\mathcal{P}_d}
    |p_P-p_P'|
    \sum_{k=0}^{2\alpha-1}
    |p_P|^{2\alpha-1-k}|p_P'|^k .
\end{align}
For each term in the sum over $k$, the Cauchy-Schwarz inequality gives 
\begin{align}
    &\frac{1}{d}
    \sum_{P\in\mathcal{P}_d}
    |p_P-p_P'| |p_P|^{2\alpha-1-k}|p_P'|^k
    \nonumber\\
    &\le
    \left[\frac{1}{d}\sum_{P\in\mathcal{P}_d}(p_P-p_P')^2
    \right]^{1/2}
    \left[\frac{1}{d}\sum_{P\in\mathcal{P}_d}|p_P|^{2(2\alpha-1-k)}|p_P'|^{2k}
    \right]^{1/2}.
\end{align}
where the second factor is bounded by 1. 
If $k=0$, then we have 
$|p_P|^{2(2\alpha-1)}\le p_P^2,$
while if $k=2\alpha-1$, then
$|p_P'|^{2(2\alpha-1)}\le(p_P')^2$.
For $0<k<2\alpha-1$, both powers are nonzero, and since $|p_P|,|p_P'|\le1$, the product is bounded by either $p_P^2$ or $(p_P')^2$. 
For a pure state, 
\begin{equation}
    \frac{1}{d}
    \sum_{P\in\mathcal{P}_d}p_P^2
    =\mathrm{tr}\rho^2=1,
    \quad
    \frac{1}{d}
    \sum_{P\in\mathcal{P}_d}(p_P')^2
    =\mathrm{tr}\rho_{\mathrm{trunc}}^2
    =1.
\end{equation}
Thus
\begin{equation}
    \left[
    \frac{1}{d}
    \sum_{P\in\mathcal{P}_d}
    |p_P|^{2(2\alpha-1-k)}
    |p_P'|^{2k}
    \right]^{1/2}
    \le1.
\end{equation}

It remains to bound the first factor. 
Using the Pauli expansion
\begin{equation}
    \rho-\rho_{\mathrm{trunc}}
    =\frac{1}{d}
    \sum_{P\in\mathcal{P}_d}
    (p_P-p_P')P,
\end{equation}
and the orthogonality relation $\mathrm{tr}(PP')=d\,\delta_{P,P'}$,
we find 
\begin{equation}
    \frac{1}{d}
    \sum_{P\in\mathcal{P}_d}
    (p_P-p_P')^2
    =\mathrm{tr}
    \left[(\rho-\rho_{\mathrm{trunc}})^2
    \right].
\end{equation}
For two pure states this becomes
\begin{align}
    \mathrm{tr}
    \left[
    (\rho-\rho_{\mathrm{trunc}})^2
    \right]
    &=
    2\left[1-\mathrm{tr}(\rho\rho_{\mathrm{trunc}})\right]
    \nonumber\\
    &=
    2\left(
    1-\left|\braket{\tilde{\psi}|\tilde{\psi}_{\mathrm{trunc}}}
    \right|^2\right)
    =2\eta .
\end{align}
Therefore,
\begin{equation}
    \left[
    \frac{1}{d}
    \sum_{P\in\mathcal{P}_d}
    (p_P-p_P')^2
    \right]^{1/2}
    =
    \sqrt{2\eta}.
\end{equation}
Since there are \(2\alpha\) terms in the expansion of
\(x^{2\alpha}-y^{2\alpha}\), we obtain
\begin{equation}
    \left|\zeta_\alpha(\ket{\tilde{\psi}})
    -\zeta_\alpha(\ket{\tilde{\psi}_{\mathrm{trunc}}})\right|
    \le 2\alpha\sqrt{2\eta}. \label{eqb18}
\end{equation}

Finally, we translate this bound into an error bound for the SRE we considered. 
Denoting 
\begin{equation}
    \delta_\alpha
    =\left|\zeta_\alpha(\ket{\tilde{\psi}})-
    \zeta_\alpha(\ket{\tilde{\psi}_{\mathrm{trunc}}})
    \right|, 
\end{equation}
and from the above result, we have 
$\delta_\alpha \le C_\alpha\sqrt{\eta}$. 
Since 
\begin{equation}
    M_\alpha(\ket{\psi})
    =\frac{1}{1-\alpha}\log_2\zeta_\alpha(\ket{\psi}),
\end{equation}
we have 
\begin{align}
    \left|
    M_\alpha(\ket{\tilde{\psi}})-M_\alpha(\ket{\tilde{\psi}_{\mathrm{trunc}}})
    \right|=\frac{1}{|\alpha-1|}
    \left|\log_2\frac{
    \zeta_\alpha(\ket{\tilde{\psi}})
    }{\zeta_\alpha(\ket{\tilde{\psi}_{\mathrm{trunc}}})}
    \right|.
\end{align}
If $\delta_\alpha\ll
\zeta_\alpha(\ket{\tilde{\psi}})$, then expanding the logarithm gives 
\begin{equation}
    \left|
    M_\alpha(\ket{\tilde{\psi}})
    -
    M_\alpha(\ket{\tilde{\psi}_{\mathrm{trunc}}})
    \right|
    \lesssim
    \frac{1}{|\alpha-1|\ln2}
    \frac{\delta_\alpha}
    {\zeta_\alpha(\ket{\tilde{\psi}})}.
\end{equation}
Using $\delta_\alpha\le C_\alpha\sqrt{\eta}$, we obtain
\begin{equation}
    \left|
    M_\alpha(\ket{\tilde{\psi}})
    -
    M_\alpha(\ket{\tilde{\psi}_{\mathrm{trunc}}})
    \right|
    \lesssim
    \frac{C_\alpha}{|\alpha-1|\ln2}
    \frac{\sqrt{\eta}}
    {\zeta_\alpha(\ket{\tilde{\psi}})}.
\end{equation}
This proves Eq.~\eqref{eq:M_trunc_bound_main}, where we use $C_{\alpha}=2\alpha\sqrt{2}$ from Eq.~\eqref{eqb18}. 
Therefore, once the accumulated discarded weight \(\eta\) is monitored and kept sufficiently small, truncating the tail of a rapidly decaying entanglement spectrum gives a controlled approximation to the nonlocal SRE. 

\section{Nonlocal SRE in Haar random states}\label{app:nonlocal_haar}
We now derive the Haar-random average of the conjectured nonlocal SRE in
the large Hilbert-space dimension limit. Consider a bipartite Haar-random
pure state with subsystem dimensions $d_A=d=2^n$ and $d_B$. We denote $\gamma=d_A/d_B$ and assume $\gamma\in(0,1]$. 
For a fixed Schmidt spectrum $\boldsymbol{\lambda}=\{\lambda_x\}_{x\in\mathbb F_2^n}$
the conjectured reference state expression gives
\begin{eqnarray}
    C(\boldsymbol{\lambda}) &=& \sum_{u,v,w,z\in\mathbb{F}_2^n}
    \sqrt{\lambda_{u}\lambda_{u\oplus v}\lambda_{u\oplus w}\lambda_{u\oplus z}\lambda_{u\oplus v\oplus w}}\times\nonumber\\
    &{}&\sqrt{\lambda_{u\oplus v\oplus z}\lambda_{u\oplus w\oplus z}\lambda_{u\oplus v\oplus w\oplus z}}\\
    &=&\sum_{u,v,w,z\in\mathbb{F}_2^n}\prod_{\omega\in\{0,1\}^3}\sqrt{\lambda_{u\oplus \omega_1 v\oplus\omega_2 w\oplus \omega_3 z}}
    \label{discrete_cube}
\end{eqnarray}
where $C(\boldsymbol{\lambda})=2^{-M_2^{\mathrm{NL}}(\boldsymbol{\lambda})}$ and we replace $i_1,i_2,i_3,i_4$ with 
$u,u\oplus v, u\oplus w, u\oplus z$. 
Here $\oplus$ denotes bitwise addition in $\mathbb F_2^n$.

For a Haar-random state, the Schmidt values are distributed according
to the induced Wishart ensemble. In the large-dimension limit with fixed
ratio $\gamma$, the rescaled eigenvalues $x_i=d\lambda_i$ 
converge to the Marchenko-Pastur distribution
\begin{equation}
    \rho_\gamma(x)=
    \frac{\sqrt{(x_+-x)(x-x_-)}}{2\pi\gamma x}, 
\end{equation}
with $x_\pm=(1\pm\sqrt{\gamma})^2$. 
After arranging the Schmidt eigenvalues in descending order, we may write
\begin{equation}
    \lambda_u
    \simeq
    \frac{1}{d}g_{\gamma}^2 \left(\frac{u}{d}\right),
    \label{lambda_quantile}
\end{equation}
where $g_\gamma^2$ is the descending quantile function of the
Marchenko-Pastur distribution, i.e., it satisfies
\begin{equation}
    \int_{g_\gamma^2(s)}^{x_+}
    \rho_\gamma(x)\,\mathrm{d}x=s, 
    \label{quantile_def}
\end{equation}
for $s\in[0, 1]$. 

We now identify the discrete Boolean label $u\in\mathbb F_2^n$ with the
dyadic point $u/d\in[0,1]$. In this notation, the bitwise XOR operation
on $\mathbb F_2^n$ converges to dyadic bitwise addition on $[0,1]$, which
we denote by the same symbol $\oplus$. Substituting
Eq.~\eqref{lambda_quantile} into Eq.~\eqref{discrete_cube}, each square-root
factor contributes
\begin{equation}
    \sqrt{\lambda_{u\oplus \omega_1 v\oplus \omega_2 w\oplus \omega_3 z}}
    \simeq
    \frac{1}{\sqrt{d}}\,
    g_\gamma\!\left(
    \frac{u\oplus \omega_1 v\oplus \omega_2 w\oplus \omega_3 z}{d}
    \right), 
\end{equation}
and we can obtain
\begin{equation}
    C(\boldsymbol{\lambda})
    =\frac{1}{d^4}\sum_{u,v,w,z} 
    \prod_{\omega\in\{0,1\}^3} g_{\gamma}\left(
    \frac{u\oplus \omega_1 v\oplus \omega_2 w\oplus \omega_3 z}{d}
    \right), 
\end{equation}
in the large-$d$ limit, the discrete sum
becomes the dyadic integral $C(\boldsymbol{\lambda})\to C_{\gamma}$, 
\begin{equation}
    C_\gamma
    =\int_{[0,1]^4}
    \prod_{\omega\in\{0,1\}^3}
    g_\gamma\left(
    u\oplus \omega_1 v\oplus \omega_2 w\oplus \omega_3 z
    \right)
    \mathrm{d}u\mathrm{d}v\mathrm{d}w\mathrm{d}z. 
    \label{Cgamma2}
\end{equation}
Here $\omega_j v$ is ordinary multiplication by
$\omega_j\in\{0,1\}$, while all additions between the selected increments
are dyadic bitwise additions.

\section{Relation between FNL and NL}\label{app4}
For a pure fermionic Gaussian state $\ket{\psi_G}$ and a bipartition $A\cup B$, the fermionic nonlocal magic (FNL) is defined as
\begin{equation}
    M_{\alpha}^{\mathrm{FNL}}:=\min_{U_{O_A},U_{O_B}}M_{\alpha}\left[
    T_{\pi_{AB}}(U_{O_A}\otimes U_{O_B})T_{\pi_{AB}}^{\dagger}\ket{\psi_G}\right],
\end{equation}
where the minimization is restricted to local fermionic Gaussian unitaries induced by 
$\mathrm{O}(2\ell)\oplus \mathrm{O}(2(L-\ell))$, with $L$ the total system size and $\ell=|A|$ the subsystem size~\cite{Daniele2026,Mario2026}. 
For fermionic Gaussian states, this restricted minimization admits a closed-form expression,
\begin{equation}
    M_{\alpha}^{\mathrm{FNL}}(\ket{\psi_G})=\sum_{i=1}^{\ell}m_{\alpha}(a_i^2),
\end{equation}
where
\begin{equation}
    m_{\alpha}(x)=\frac{1}{1-\alpha}\log_2\left[\frac{(1-x)^{\alpha}+1+x^{\alpha}}{2}\right].
\end{equation}
Here $\{a_i\}_{i=1,\dots,\ell}$ are the positive eigenvalues of $i\Gamma_A$, where $\Gamma_A$ is the $2\ell\times 2\ell$ 
reduced Majorana covariance matrix of subsystem $A$. 
In particular, for $\alpha=2$, one has $m_2(x)=-\log_2(1-x+x^2)$.

We note that the eigenvalues $\{a_i\}$ of the reduced covariance matrix are related with the many-body entanglement spectrum. 
The reduced density matrix of a fermionic Gaussian state can be written as 
\begin{equation}
    \rho_A=\frac{e^{-H_{\mathrm{ent}}}}{Z_{\mathrm{ent}}},
    \qquad H_{\mathrm{ent}}=\sum_i \epsilon_i^{\mathrm{ent}} f_i^\dagger f_i,
\end{equation}
with $a_i=\tanh(\epsilon_i^{\mathrm{ent}}/2)$. 
Equivalently, each entanglement mode contributes two probabilities
\begin{equation}
    p_i(0)=\frac{1+a_i}{2},
    \qquad
    p_i(1)=\frac{1-a_i}{2}.
\end{equation}
The full many-body entanglement spectrum is therefore
\begin{equation}
    \lambda_{\mathbf n}
    =
    \prod_{i=1}^{\ell}
    \left(\frac{1+a_i}{2}\right)^{1-n_i}
    \left(\frac{1-a_i}{2}\right)^{n_i}.
\end{equation}
with $n=(n_1,\dots,n_\ell)\in\{0,1\}^{\ell}$. 
Thus, the covariance-matrix eigenvalues $\{a_i\}$ determine the entanglement spectrum. 

Therefore, the relation between FNL and our nonlocal magic resource (NL) is then transparent.  
Since the spectrum $\lambda_{\boldsymbol{n}}$ is not sorted, so for the same entanglement spectrum, we have  
\begin{equation}
    M_{\alpha}^{\mathrm{FNL}}\geq M_{\alpha}^{\mathrm{NL,ref}}.
\end{equation}
Importantly, this inequality does not rely on our conjecture and holds generally. 
In this sense, FNL provides a natural, calculation-friendly upper bound on the nonlocal SRE. 
The difference between the two quantities comes from the ordering structure of the entanglement spectrum. 
The closed-form expression for FNL corresponds to the Gaussian, or BCS, canonical ordering of the spectrum, 
in which the many-body probabilities retain their product structure in the fermionic occupation basis. 
By contrast, our nonlocal SRE is obtained by sorting the same set of Schmidt coefficients in descending order before assigning them to the computational basis. 
Therefore, FNL and NL are computed from the same entanglement spectrum but with different binary labelings. 

\section{Further discussion on the PXP eigenstate properties}\label{app:pxp}
\begin{figure}
    \includegraphics[width=\linewidth]{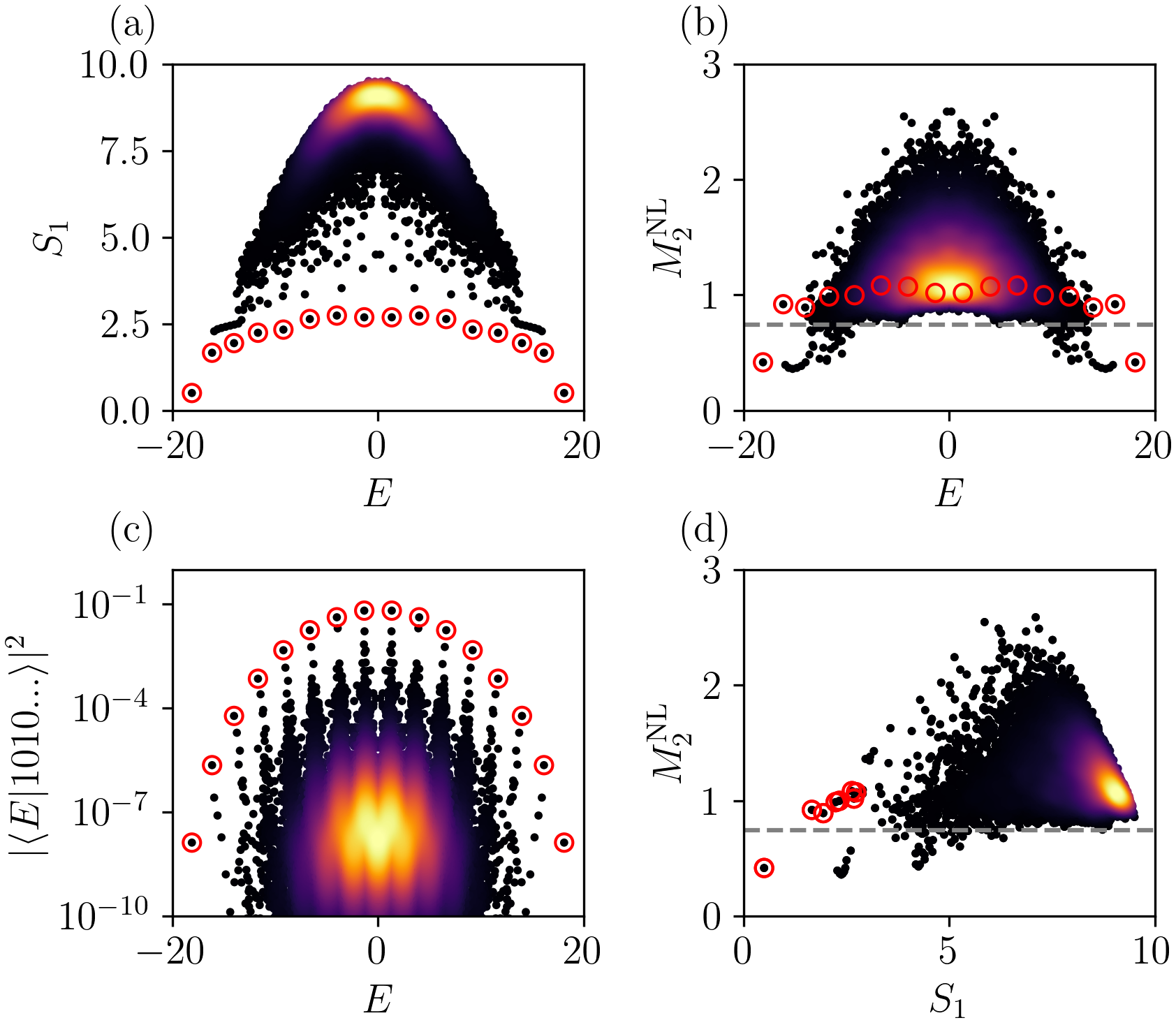}
    \caption{
    Eigenstate properties of the PXP model. 
    We diagonalize the $N=30$ PXP Hamiltonian in the zero-momentum and inversion-symmetric sector under periodic boundary conditions. 
    (a) Bipartite entanglement entropy $S_1$ as a function of eigenenergy $E$. 
    (b) Nonlocal SRE $M_2^{\mathrm{NL}}$ of the eigenstates, where the gray dashed line indicates the Haar-random average. 
    (c) Overlap of the N\'eel state $\ket{1010\cdots}$ with the energy eigenstates. 
    (d) Distribution of eigenstates in the $(S_1,M_2^{\mathrm{NL}})$ plane. 
    Scarred eigenstates are highlighted by red circles. 
    Brighter colors indicate a higher density of eigenstates.
    \label{fig9}
    }
\end{figure}

Beyond the dynamical analysis, we also investigate the eigenstate properties of the PXP model, as shown in Fig.~\ref{fig9}. 
Figure~\ref{fig9}(b) shows that the majority of eigenstates have nonlocal SRE close to the Haar-random average. 
This is consistent with the eigenstate thermalization hypothesis~\cite{Deutsch2018,Rigol2008,Srednicki1994}, since thermal eigenstates possess nearly maximal entanglement entropy and an approximately flat many-body entanglement spectrum, causing their nonlocal SRE to approach the Haar value. 
Only a small fraction of eigenstates exhibit substantially larger nonlocal SRE, indicating more structured entanglement spectra.

The scarred eigenstates behave differently. 
As shown in Fig.~\ref{fig9}(a), they possess anomalously low entanglement entropy, in clear violation of the thermal expectation. 
Nevertheless, their nonlocal SRE remains comparable to, and slightly above, the Haar average rather than being strongly suppressed. 
Consequently, in the $(S_1,M_2^{\mathrm{NL}})$ plane shown in Fig.~\ref{fig9}(d), the scarred eigenstates form a distinct branch separated from the thermal cloud. 
This indicates that nonlocal SRE is not determined solely by the amount of entanglement, but is also sensitive to the detailed organization of the ordered Schmidt spectrum. 
While thermal eigenstates approach the Haar value through the flattening of their entanglement spectra, scarred eigenstates retain a comparable nonlocal SRE despite their much lower entanglement entropy, highlighting that nonlocal SRE probes spectral structures beyond conventional entanglement measures. 
The enhanced nonlocal SRE observed during scar dynamics is therefore not a direct consequence of unusually large nonlocal SRE in the scarred eigenstates themselves, but instead originates from their coherent superposition and long-lived interference during the nonequilibrium evolution.

\nocite{*}
\bibliography{ref}
\end{document}